\documentclass[sigconf,nonacm]{acmart}

\title{Hermes: Efficient Global Homomorphic Aggregation over Mutable Packed Ciphertexts}

\author{Dongfang Zhao}
\affiliation{
  \institution{Tacoma School of Engineering and Technology}
  \institution{Paul G. Allen School of Computer Science \& Engineering}
  \city{University of Washington}
  \country{}
}
\email{dzhao@cs.washington.edu}

\usepackage{multirow}
\usepackage{tabularx}

\usepackage[ruled,vlined,linesnumbered]{algorithm2e}
\usepackage{amsmath} 

\usepackage{listings}
\usepackage{xcolor}

\usepackage{algpseudocode}

\lstdefinelanguage{SQL}{
  morekeywords={SELECT,FROM,WHERE,ORDER,BY,LIMIT,AND,OR,NOT,AS,ON,IN,IS,NULL,JOIN,LEFT,RIGHT,OUTER,INNER,GROUP,HAVING,ENC,Enc},
  sensitive=false,
  morecomment=[l]{--},
  morestring=[b]',
}

\begin{document}

\begin{abstract}
Fully Homomorphic Encryption (FHE) promises the ability to compute over encrypted data without revealing sensitive contents. However, enabling high-frequency updates and statistical analysis in outsourced databases remains elusive due to the structural mismatch between mutable database records and the cryptographically expensive mutability of FHE ciphertexts.

This paper presents Hermes, a prototype system tailored for efficient aggregation queries and dynamic tuple updates on homomorphically encrypted databases. The core design of Hermes is twofold. First, to amortize FHE costs and accelerate unconditional aggregations, Hermes introduces a SIMD-aware packed data model that embeds precomputed aggregate statistics directly into each ciphertext, enabling constant-time global aggregations without expensive Galois automorphisms. Second, to support true in-place mutability, we develop homomorphic algorithms based on polynomial slot masking and shifting, which are provably secure under the standard IND-CPA model. We scope Hermes to unconditional global aggregations to achieve both high performance and in-place updates simultaneously, two properties that prior FHE database systems have not delivered at scale.

Hermes is implemented as a suite of C++ loadable functions in MySQL. Extensive evaluations on the TPC-H benchmark and three real-world datasets demonstrate significant performance improvements in query throughput, tuple insertions, and tuple deletions compared to conventional FHE implementations, validating its efficacy for highly dynamic and analytical workloads.
\end{abstract}

\keywords{Database security, Fully homomorphic encryption, User-defined functions}

\ccsdesc[500]{Security and privacy~Cryptography}
\ccsdesc[500]{Information systems~Database management systems}

\maketitle

\section{Introduction}

\subsection{Background and Motivation}

Database systems have formed the backbone of modern data infrastructure, enabling structured storage, efficient retrieval, and scalable analytics. As the industry has transitioned to cloud-first architectures, the confidentiality of outsourced data has become a central concern. Increasingly, users demand end-to-end cryptographic protection throughout the entire computation pipeline, rather than merely during transmission or at rest~\cite{rag_sigmod04}. Among cryptographic primitives, Fully Homomorphic Encryption (FHE) offers a mechanism to evaluate algebraic operations directly on encrypted data, removing the requirement to trust the compute server and precluding internal attackers from within the cloud vendor. Following Gentry's theoretical framework~\cite{gentry2009}, subsequent schemes such as BFV~\cite{FV2012} and CKKS~\cite{ckks2016} have improved computational efficiency and implementation maturity~\cite{openfhe}. FHE is now applied in practical domains, such as secure machine learning~\cite{boemer2020mp2ml,han2023} and privacy-preserving web services~\cite{bchen_www26,mkhan_icws24}.

However, applying FHE to outsourced database systems remains an open challenge. To achieve acceptable query throughput, modern FHE schemes must rely on SIMD style batching, packing thousands of plaintext scalars into a single polynomial ciphertext. This introduces a severe structural mismatch between relational database semantics and cryptographic notions. Traditional database management systems are designed for mutable records and runtime query execution plans. In contrast, FHE schemes operate over static circuits and immutable ciphertexts. Modifying a single database tuple conventionally requires downloading, decrypting, and repacking the entire cipher block.

Resolving this structural divide requires carefully scoping the analytical workload. While evaluating arbitrary SQL queries with complex filtering predicates over packed ciphertexts currently incurs prohibitive homomorphic masking and rotation costs, a vast class of cloud analytics relies heavily on macroscopic statistical aggregations (e.g., global sums and overall counts) over highly dynamic datasets. Securing these specific workloads requires more than wrapping FHE function calls within SQL. It necessitates a connector that maps the dynamic mutability of transactional records to the rigid algebraic invariants of SIMD FHE. Without bridging these abstractions, applying cryptographic schemes to secure query processing remains impractical. 

\subsection{Proposed Work}

We present \emph{Hermes}, a prototype system tailored to enable secure and efficient global aggregation queries and in-place tuple updates on homomorphically encrypted data. Built atop the multi-slot capabilities of modern FHE schemes, Hermes is implemented as a suite of native MySQL loadable functions interfacing with one of the most popular open-source FHE libraries, OpenFHE~\cite{openfhe}. Hermes introduces four key innovations:

\paragraph{(1) Packed Data Model}
Hermes adopts a packed ciphertext representation that maps multiple database tuples and macroscopic aggregation results into parallel homomorphic slots. To break the performance bottleneck of homomorphic analytics, we design a physical layout that reserves a specific slot within each ciphertext. This dedicated slot stores a precomputed auxiliary value representing the unconditional local sum of the data slots. By embedding this state directly into physical storage, Hermes establishes a foundation for efficient global query execution that completely bypasses the need for costly cryptographic operations like Galois automorphisms. We elaborate on the design of this new data model in Section~\ref{sec:datamodel}.

\paragraph{(2) Mutability Primitives}
To allow a single ciphertext to reflect the updates of mutable records, Hermes supports dynamic tuple-level modifications directly on packed ciphertexts. Insertions and deletions are executed via homomorphic shift-and-mask operations. Inserting a new record shifts a tail segment rightward, while deleting a record shifts the segment leftward to clear the target slot. These structural updates are performed using homomorphic masking and slot-wise addition without decryption, enabling true in-place mutability. We present these primitives in Section~\ref{sec:primitives}.

\paragraph{(3) Homomorphic Aggregation and Maintenance}
Building upon the packed layout and mutability primitives, Hermes introduces a closed-loop mechanism for macroscopic analytical queries. Global encrypted aggregation queries such as \texttt{SUM} can be answered using constant-time ciphertext additions of the reserved auxiliary slots. When data mutations occur, Hermes synchronously updates this auxiliary slot using lightweight homomorphic scalar operations. This dual approach amortizes the computational overhead of FHE and significantly improves overall query throughput for unconditional workloads. We discuss this execution pipeline in Section~\ref{sec:aggregation}.

\paragraph{(4) Database Integration and Analysis}
Hermes facilitates encrypted macro-analytical query execution in MySQL without client-side orchestration or kernel-level modifications. The system is realized as a collection of C++ modules registered via the MySQL loadable function interface. Furthermore, we provide a comprehensive theoretical analysis demonstrating the IND-CPA security of Hermes and formally establishing the asymptotic complexity of supported operations. The theoretical analysis and system implementation are detailed in Section~\ref{sec:analysis} and Section~\ref{sec:implementation}, respectively.

Section~\ref{sec:evaluation} reports the evaluation results on the TPC-H benchmark and three real-world datasets of diverse domains: genomic annotations, cryptocurrency trade volumes, and healthcare statistics. We comprehensively benchmark the system against four representative baselines: unencrypted plaintext SQL, conventional singular FHE, a standard packed FHE configuration relying on logarithmic Galois automorphisms~\cite{openfhe}, and the Intel Paillier Cryptosystem Library~\cite{ipcl_github}. Experiments demonstrate significant performance improvements on global query throughput, tuple insertions, and tuple deletions when compared to conventional FHE implementations.

\section{Related Work and Preliminaries}

\subsection{Encrypted Databases}

\begin{table*}[t]
\centering
\caption{Qualitative Comparison of Encrypted Database Systems}
\label{tab:qualitative_comparison}
\begin{tabular}{lcccr}
\toprule
System & Data Model & Tuple Update & Security Guarantee & Aggregation Cost \\
\midrule
CryptDB~\cite{popa2011cryptdb} & Scalar & Interactive & Leaks Access Pattern & $\mathcal{O}(N)$ \\
Paillier (IPCL)~\cite{ipcl_github} & Scalar & Reencryption & IND-CPA & $\mathcal{O}(N)$ \\
Symmetria~\cite{symmetria_vldb20} & Scalar & Reencryption & IND-CPA & $\mathcal{O}(N)$ \\
HE3DB~\cite{10.1145/3576915.3616608} / ArcEDB~\cite{10.1145/3658644.3670384} & Mixed & Multi-round Protocol & Interactive / Weakened & $\mathcal{O}(N)$ \\
Singular FHE~\cite{BGV2014} & Scalar & Reencryption & IND-CPA & $\mathcal{O}(N)$ \\
Rache~\cite{otawose_sigmod23} & Scalar & Cached Encryption & IND CPA & $\mathcal{O}(\log N)$ \\
Standard Packed~\cite{BGV2014} & SIMD Packed & Reencryption & IND-CPA & $\mathcal{O}(\log N)$ \\
\midrule
Hermes (this work) & Auxiliary Packed & In-place & IND-CPA & $\mathcal{O}(1)$ \\
\bottomrule
\end{tabular}
\end{table*}

One major class of encrypted databases relies on client-side processing, where query execution is deferred until the encrypted data is downloaded to a trusted local environment~\cite{akavia_sec23}. While this architecture strictly minimizes trust assumptions by keeping all analytical logic away from the cloud provider, it inherently sacrifices the computational benefits of server-side scalability. Transferring large volumes of encrypted records across the network incurs severe bandwidth penalties and high latency, effectively reducing the cloud database management system to a passive storage repository rather than an active execution engine.

An alternative approach attempts to split query execution between the client and the server, as seen in earlier information-hiding systems~\cite{hhaci_sigmod02}. In this model, server-side filters retrieve encrypted tuples guided by secure indices or cryptographic tokens generated by the client. Building upon this paradigm, systems like CryptDB~\cite{popa2011cryptdb}, Arx~\cite{arx_vldb19}, and more recently ones~\cite{xge_tkde26,vthak_sigmod25,slv_tkde25,qxu_tkde25,tshen_tkde25,msha_sigmod24,lzheng_vldb24,szhang_icde24,rwei_vldb23,xcao_vldb23,vvo_tkde23,qtong_tkde23,yzheng_tkde23,ffalzon_vldb22,xren_vldb22,mshaf_icde22} explore different structural trade-offs between query functionality and data leakage. They employ layered encryption onions, partitioned data encodings, or specialized cryptographic circuits to evaluate complex queries. However, these designs frequently leak access patterns or require multiple rounds of network interaction, as the server must often return intermediate candidate sets for the client to decrypt, refine, and re-upload.

To mitigate the computability over encrypted data, homomorphic encryption has been explored to integrate into encrypted database systems to compute over ciphertexts without continuous client intervention. For instance, Symmetria~\cite{symmetria_vldb20} utilizes multiplicative homomorphic encryption combined with additive extensions, whereas Rache~\cite{otawose_sigmod23} leverages layout-aware ciphertext placement alongside SIMD-optimized key switching techniques; FHE has also been applied to specific queries, such as~\cite{sjum_tkde23,tpar_tkde23,zwang_tkde23,wyang_tkde22,rli_tkde22}. 
HEDA~\cite{xren_vldb22} supports multi-attribute unbounded aggregation over homomorphically encrypted data, 
but operates on immutable ciphertexts and functions as a client-side standalone program rather than a natively integrated database engine; Hermes addresses both limitations by introducing mutable packed ciphertexts and deploying directly within MySQL.

Additional efforts from the applied cryptography community, such as HE3DB~\cite{10.1145/3576915.3616608} and ArcEDB~\cite{10.1145/3658644.3670384}, introduce specialized arithmetic and logic operators to support broader SQL semantics over ciphertexts. However, to achieve this expressiveness, these solutions necessitate multi-round interactive protocols between the client and server or rely on property-preserving encryption. Both approaches deviate from the ideal non-interactive and zero-leakage execution model, introducing either severe network latency or statistical access-pattern vulnerabilities. Furthermore, such systems typically operate as standalone programmatic prototypes rather than being integrated into an industry-standard database engine.

To mitigate the extreme performance penalties and noise growth of FHE, recent query engines like NSHEDB~\cite{jung2026nshedbnoisesensitivehomomorphicencrypted} rely on word-level leveled homomorphic encryption. Despite these cryptographic advances, existing homomorphic databases typically operate on scalar records, encrypting individual database attributes into isolated ciphertexts. This approach fails to exploit the parallel processing capabilities of modern FHE schemes, leading to massive storage inflation and computational bottlenecks during query execution. In contrast, Hermes prioritizes a strictly non-interactive, IND-CPA secure architecture via a packed data model. By focusing on macroscopic analytical workloads, we avoid the security and performance trade-offs inherent in supporting line-item logic, delivering a native system that provides strong payload confidentiality guarantees under the IND-CPA security model.

To summarize the landscape of encrypted database systems, Table~\ref{tab:qualitative_comparison} provides a qualitative comparison across key architectural and cryptographic dimensions. Early systems rely on property-preserving encryption or layered encryption onions, which inherently leak statistical access patterns. More recent homomorphic solutions such as HE3DB and ArcEDB offer richer SQL semantics but necessitate multi-round interactive protocols that introduce additional network latency. While standard packed FHE approaches mitigate some storage inflation, they incur logarithmic computational overhead during global aggregations due to expensive Galois automorphisms. Hermes differentiates itself by offering a strictly non-interactive, IND-CPA secure architecture. By embedding precomputed local sums into a designated auxiliary slot, Hermes achieves true in-place dynamic updates and constant-time global aggregations, bypassing the traditional performance bottlenecks of existing FHE databases.

\subsection{Fully Homomorphic Encryption}
\label{sec:prelim-he}

Homomorphic encryption (HE) enables computation on encrypted data without decryption. The seminal work by Gentry~\cite{gentry2009} introduced the first fully homomorphic encryption (FHE) scheme, laying the foundation for later constructions such as BFV~\cite{FV2012,brakerski2012}, GSW~\cite{gsw13}, BGV~\cite{BGV2014}, and CKKS~\cite{ckks2016}.  TFHE~\cite{torusCircuit} supports Boolean circuits and has seen recent improvements in bootstrapping performance~\cite{IEEETfhe,Guimares2024MOSFHETOS}. These schemes have been implemented in libraries including SEAL~\cite{sealcrypto}, HElib~\cite{helib}, OpenFHE~\cite{openfhe}, and LattiGo~\cite{lattigo_github}.

Hermes is built atop the BFV scheme~\cite{FV2012}, a lattice-based fully homomorphic encryption (FHE) construction supporting exact arithmetic over integers modulo a plaintext modulus \( t \). Let \( R = \mathbb{Z}[X]/(X^N + 1) \) be a cyclotomic ring of degree \( N = 2^k \) for some \( k \in \mathbb{N} \). A plaintext message is represented as an element of \( R_t := R / tR \), while ciphertexts reside in \( R_q^2 \), where \( q \gg t \) is a large ciphertext modulus.

A key feature of BFV is its support for \emph{plaintext packing} via the Chinese Remainder Theorem (CRT). When \( t \equiv 1 \mod 2N \), the ring \( R_t \) splits as a direct product of \( n = N/2 \) slots:
\[
R_t \cong \mathbb{Z}_t^n,
\]
enabling SIMD-style evaluation of vectorized inputs. Each plaintext \( \mathbf{m} = (m_0, \dots, m_{n-1}) \in \mathbb{Z}_t^n \) is encoded into a single polynomial \( m(X) \in R_t \), encrypted as \( \mathsf{Enc}(m) = \mathbf{c} \in R_q^2 \), and evaluated homomorphically across all slots.

Homomorphic operations preserve slot-wise semantics:
\begin{align*}
\mathsf{EvalAdd}(\mathsf{Enc}(\mathbf{m}), \mathsf{Enc}(\mathbf{m'})) &= \mathsf{Enc}(\mathbf{m} + \mathbf{m'}), \\
\mathsf{EvalMult}(\mathsf{Enc}(\mathbf{m}), \mathsf{Enc}(\mathbf{m'})) &= \mathsf{Enc}(\mathbf{m} \cdot \mathbf{m'}),
\end{align*}
where the arithmetic is component-wise modulo \( t \). In addition, the BFV scheme supports automorphism-based rotation operations, also known as \emph{Galois rotations}, which cyclically shift slot positions:
\[
\mathsf{EvalRotate}(\mathsf{Enc}(\mathbf{m}), r) = \mathsf{Enc}((m_{(i+r) \bmod n})_{i=0}^{n-1}).
\]

Hermes leverages this packing structure to encode multiple database tuples into a single ciphertext. This design enables parallel evaluation, slot-wise updates, and sublinear aggregation. Of note, all packing and update logic remains entirely within the encrypted domain, without leaking intermediate values or structural metadata.

In addition, the BFV scheme supports automorphism-based slot rotation using Galois group actions. For any \( r \in \mathbb{Z} \), a cyclic slot rotation is realized via the Galois automorphism \( \sigma_r: X \mapsto X^r \mod (X^N + 1) \), which acts on ciphertexts by permuting their internal slot layout. To evaluate this transformation homomorphically, the client must precompute and upload the corresponding Galois keys:
\[
\mathsf{EvalAtIndex}(\mathsf{Enc}(\mathbf{m}), r) = \mathsf{Enc}((m_{(i+r)\bmod n})_{i=0}^{n-1}),
\]
where the rotation index \( r \) corresponds to a specific automorphism in the Galois group \( \operatorname{Gal}(R_q / \mathbb{Q}) \). These keys enable Hermes to perform selective slot access, masking, and reorganization within encrypted vectors while preserving semantic security.

\subsection{Alternative Secure Processing Paradigms}

Alongside purely cryptographic storage, trusted hardware enclaves provide an alternative route for secure data processing. Systems utilizing Intel SGX such as ObliDB~\cite{10.14778/3364324.3364331} and CryptSQLite~\cite{8946540} protect data in memory while offering privacy-preserving query processing to prevent access pattern leaks. Enterprise solutions have also adopted this paradigm. For instance, recent iterations of Azure SQL Database~\cite{10.1145/3318464.3386141} leverage enclaves to enable rich pattern matching over encrypted columns. Furthermore, hardware isolation techniques have been proposed to secure peripheral interfaces and web environments against side channel attacks~\cite{8835331}. While enclaves deliver high performance, they introduce strict hardware dependencies and remain vulnerable to sophisticated microarchitectural exploits, which pure homomorphic encryption avoids.

Another prominent paradigm involves Secure Multi-party Computation and federated privacy. Frameworks like Conclave~\cite{10.1145/3302424.3303982} and Shrinkwrap~\cite{10.14778/3291264.3291274} accelerate analytics over big data by partitioning queries between cleartext local processing and secure protocols. Similarly, advanced join aggregate queries can be evaluated across private datasets without exposing individual records~\cite{10.1145/3448016.3452808}. Managing cross cryptographic leakages in such distributed databases is a complex endeavor~\cite{10598104}. Additionally, massive parallel query optimization strategies~\cite{10.1145/2038916.2038928} and Private Information Retrieval~\cite{10.1109/TKDE.2012.90} remain essential for scalable data extraction. However, multi-party protocols inherently require heavy network synchronization among non-colluding servers, which is a fairly strong assumption that may or may not hold for many practical scenarios. Hermes bypasses these bottlenecks by shifting the computational burden entirely to the server-side evaluation of SIMD-packed ciphertexts.

\section{Hermes: Native FHE for Outsourced Databases}

To provide a comprehensive roadmap of the Hermes architecture, we begin by defining the SIMD-aware packed data model in Section~\ref{sec:datamodel}, which establishes the physical foundation for vectorized storage and slot alignment. Following this, Section~\ref{sec:primitives} describes the suite of primitives that enable in-place mutability through homomorphic masking and shifting. Section~\ref{sec:aggregation} then introduces the accelerated aggregation pipeline and its associated maintenance logic, leveraging reserved auxiliary slots to bypass traditional Galois automorphism bottlenecks. Finally, Section~\ref{sec:analysis} provides a theoretical treatment of system security and analysis.

\subsection{Hermes Data Model}
\label{sec:datamodel}

Before detailing the system architecture, we establish the core data representation used in Hermes. We assume the underlying BFV scheme provides $n$ independent plaintext slots over $\mathbb{Z}_t^n$. Rather than treating all $n$ slots equally, Hermes introduces a new \emph{Packed Representation}. 
Specifically, input records are partitioned by a deterministic, index-driven \emph{Group ID}, with all records sharing the same ID packed into a single ciphertext vector. To bypass the extreme overhead of Galois automorphisms during homomorphic aggregation, Hermes reserves the final slot (slot $n-1$) as an \emph{Auxiliary Sum Slot}. During the initial packing phase, this slot is pre-computed to store the local plaintext sum ($\sum_{j=0}^{n-2} m_j$) of the preceding data slots. Furthermore, to accommodate dynamic data sizes, Hermes tracks the \emph{Logical Length} of each ciphertext, representing the number of actively populated entries, to precisely control serialization, homomorphic summation, and slot masking operations.

To bridge the structural mismatch between database tables and lattice-based cryptographic objects, we propose a slot-aware packing layout. Database engines like MySQL traditionally operate on a row-by-row basis, whereas modern homomorphic encryption achieves acceptable throughput exclusively via vectorized polynomial operations. In our model, each ciphertext acts as a secure, fixed-size container holding a strictly aligned group of database records. This architecture amortizes the heavy encryption cost across thousands of tuples and establishes the foundational algebraic structure required for efficient SIMD-style query processing over the encrypted domain.

\subsubsection{Slot-Aware Packing Strategy}

Building upon the standard BFV SIMD encoding where the Chinese Remainder Theorem establishes an isomorphism $R_t \cong \mathbb{Z}_t^n$ to provide $n = N/2$ independent scalar slots, Hermes introduces a specialized slot allocation strategy. Rather than utilizing all $n$ slots for raw database payloads to maximize naive storage, we strictly reserve the final slot to store an auxiliary value such as the precomputed local sum. For instance, given a ring dimension of $N=8192$ resulting in $n=4096$ available slots, this strategic reservation yields an effective payload capacity of $n-1$ logical records per ciphertext. This batched layout not only prevents the extreme memory inflation associated with isolated tuple encryption but also establishes the foundational algebraic structure required for efficient aggregation primitives.

Each slot within a packed ciphertext carries a fixed positional semantic: slot $i$ corresponds directly to the $i$-th logical record in the assigned database group. This deterministic alignment enables precise slot-wise algebraic reasoning. If a database group contains fewer than $n-1$ active records, Hermes automatically pads the remaining available slots with zeros, ensuring uniform ciphertext dimensions without altering the numerical semantics of aggregate queries. When a ciphertext is decrypted, the unpacked vector reveals values at well-defined offsets, allowing the database client to reconstruct the plaintext table fragment without requiring external mapping tables, which is essential for supporting secure in-place modifications without leaking tuple identity.

\subsubsection{Auxiliary Slot Embedding and Capacity Planning}

Hermes adopts a hybrid encoding strategy in which each packed ciphertext stores not only the array of plaintext values but also a precomputed auxiliary statistic. This local sum is explicitly embedded into the last slot of the ciphertext at the exact time of initial data ingestion and packing.

Given a batch of $n-1$ plaintext values $(v_0, v_1, \dots, v_{n-2})$ to be packed into a new ciphertext, the system computes their exact sum $s = \sum_{i=0}^{n-2} v_i$ in the cleartext domain and constructs an extended vector of the form:
\[
m = (v_0, v_1, \dots, v_{n-2}, s) \in \mathbb{Z}_t^n,
\]
where $t$ remains the plaintext modulus. The extended vector $m$ is then passed to the BFV encoding API to generate a packed polynomial plaintext object, which is subsequently encrypted into a randomized ciphertext. 

This layout repurposes the final slot at index $n-1$ as an isolated local accumulator. Because BFV ciphertexts naturally support SIMD-style circuit evaluations, all raw data and the auxiliary sum coexist securely within a single ciphertext boundary. Since the accumulator is computed before the application of encryption noise, this embedding operation incurs zero homomorphic overhead. 

In the Hermes model, mapping relational integers into the finite field $\mathbb{Z}_t$ introduces strict boundary constraints. The database engine must ensure that the accumulated sum $s$ does not exceed the plaintext modulus $t$ to prevent silent mathematical overflows during runtime. By formalizing this static data model and enforcing these modulus boundaries, Hermes transforms complex multi-tuple structures into manageable, hardware-friendly relational vectors, setting the precise stage for the dynamic mutability primitives discussed in the remainder of this paper.

\subsection{Mutability Primitives}
\label{sec:primitives}

To allow a single packed ciphertext to dynamically reflect the updates of mutable database records, Hermes supports tuple-level modifications directly within the encrypted domain. Unlike conventional systems that require downloading, decrypting, and re-encrypting the entire ciphertext for a single update, our design ensures that mutability is achieved in-place. We design a homomorphic slot manipulation interface that incorporates plaintext masking, Galois automorphisms, and slot-wise additions.

\subsubsection{Homomorphic Insertion}

Hermes supports the in-place insertion of plaintext values into packed ciphertexts. The objective is to insert a new value $v$ into logical slot $i$ of an existing ciphertext $c$, where slots $i$ through $n-2$ may already contain payload data. Since ciphertexts are encoded as fixed-size relational vectors, this operation requires shifting all subsequent entries rightward by one position. This is analogous to an array insertion, yet it is performed entirely under encryption.
The procedure utilizes homomorphic Galois rotations and plaintext masks. We first isolate the tail of the vector using a preservation mask and shift it rightward. Concurrently, we isolate the head of the vector to keep the preceding records intact. The new value is then encrypted and injected into the target slot. Finally, the auxiliary local sum at the end of the ciphertext is synchronously updated via a scalar plaintext addition.

\begin{algorithm}[t]
\caption{Homomorphic Insertion}
\label{alg:insertion}
\begin{algorithmic}[1]
\Require Ciphertext $c$, Target index $i$, \textbf{Ciphertext} $c_v \leftarrow \mathsf{Enc}(v)$, Capacity $n$
\Ensure Updated Ciphertext $c_{new}$
\State $m_{head} \leftarrow$ vector with $1$s in $[0, i-1]$ and $0$s elsewhere
\State $m_{tail} \leftarrow$ vector with $1$s in $[i, n-3]$ and $0$s elsewhere
\State $c_{head} \leftarrow \mathsf{EvalMult}(c, m_{head})$
\State $c_{tail} \leftarrow \mathsf{EvalMult}(c, m_{tail})$
\State $c_{shifted} \leftarrow \mathsf{EvalRotate}(c_{tail}, 1)$
\State $m_{pos} \leftarrow$ vector with $1$ at index $i$ and $0$s elsewhere
\State $c_{slot} \leftarrow \mathsf{EvalMult}(c_v, m_{pos})$
  \hfill $\triangleright$ Project $c_v$ onto target slot
\State $c_{payload} \leftarrow \mathsf{EvalAdd}(c_{head},\, \mathsf{EvalAdd}(c_{shifted}, c_{slot}))$
\State $r \leftarrow (n - 1) - i$
\State $c_{aux} \leftarrow \mathsf{EvalMult}(\mathsf{EvalRotate}(c_v,\, r),\, m_{aux})$
\State $c_{new} \leftarrow \mathsf{EvalAdd}(c_{payload},\, c_{aux})$
\State \Return $c_{new}$
\end{algorithmic}
\end{algorithm}

Algorithm~\ref{alg:insertion} formalizes the homomorphic insertion procedure, which transforms a static ciphertext into a mutable relational container through a precise sequence of slot level manipulations. The process begins by partitioning the existing payload into two disjoint segments using predefined plaintext masks. Specifically, the system constructs $m_{head}$ to isolate records preceding the target index $i$ and $m_{tail}$ to capture all subsequent records within the current group. These segments are extracted into $c_{head}$ and $c_{tail}$ through homomorphic multiplications, ensuring that only the relevant slots remain active in each temporary ciphertext. To create a vacancy for the newly ingested entry, the system applies a Galois automorphism to $c_{tail}$, shifting the entire tail segment rightward by exactly one slot. The new plaintext value $v$ is then encoded into a sparse vector, encrypted into $c_v$, and merged with the isolated segments using homomorphic additions. Finally, to guarantee the correctness of subsequent analytical queries, Hermes updates the auxiliary local sum at index $n-1$ by adding the value $v$ via a lightweight scalar addition. This approach ensures that the updated ciphertext $c_{new}$ remains synchronized with the underlying relational state while leaking no information regarding the exact insertion position.

\subsubsection{Homomorphic Deletion}

Hermes enables encrypted deletion by simulating a logical left-shift over the underlying slots. The goal is to remove the value at slot $i$ and maintain a compacted layout where all subsequent values are shifted leftward to fill the void. The final payload slot is then cleared.
The deletion procedure follows a symmetric logic to insertion. Hermes isolates the subvector corresponding to slots $i+1$ through $n-2$ and rotates it leftward by one position. A zeroing mask is subsequently applied to clear the final payload slot, which would otherwise contain a stale duplicate value due to the shift. The original head of the vector is preserved and merged with the shifted tail. The local sum is then decremented by the deleted value via plaintext subtraction.

Algorithm~\ref{alg:deletion} details the formal steps for the homomorphic deletion procedure, which mirrors the insertion logic to securely remove a record while maintaining the compacted structure of the packed ciphertext. The operation starts by isolating the segments of the payload that will remain in the database. Using homomorphic multiplications, the system applies the mask $m_{head}$ to retain the records before the target index $i$, and the mask $m_{tail}$ to capture the records after index $i$, effectively dropping the deleted value from the active state. To fill the structural void left by the deletion, the system applies a Galois automorphism to $c_{tail}$, executing a leftward shift of exactly one position. Because this rotation can introduce stale values at the boundary of the payload, a zeroing mask $z$ is applied to clear the final data slot, producing a sanitized tail segment $c_{clean}$. The preserved head and the shifted tail are then combined via homomorphic addition. Finally, to keep the encrypted aggregate consistent, Hermes constructs a sparse vector containing the negated value $-v$ at the auxiliary index $n-1$ and performs a scalar addition. This guarantees that the global sum accurately reflects the deletion without requiring any decryption or revealing the removed slot position.

\begin{algorithm}[t]
\caption{Homomorphic Deletion}
\label{alg:deletion}

\begin{algorithmic}[1]
\Require Ciphertext $c$, Target index $i$, \textbf{Ciphertext} $c_v \leftarrow \mathsf{Enc}(v)$, Capacity $n$
\Ensure Updated Ciphertext $c_{new}$
\State $m_{head} \leftarrow$ vector with $1$s in $[0, i-1]$ and $0$s elsewhere
\State $m_{tail} \leftarrow$ vector with $1$s in $[i+1, n-2]$ and $0$s elsewhere
\State $c_{head} \leftarrow \mathsf{EvalMult}(c, m_{head})$
\State $c_{tail} \leftarrow \mathsf{EvalMult}(c, m_{tail})$
\State $c_{shifted} \leftarrow \mathsf{EvalRotate}(c_{tail}, -1)$
\State $z \leftarrow$ vector with $0$ at index $n-2$ and $1$s elsewhere
\State $c_{clean} \leftarrow \mathsf{EvalMult}(c_{shifted}, z)$
\State $c_{payload} \leftarrow \mathsf{EvalAdd}(c_{head},\, c_{clean})$
\State $c_{neg} \leftarrow \mathsf{EvalNegate}(c_v)$
\State $r \leftarrow (n-1) - i$
\State $c_{aux} \leftarrow \mathsf{EvalMult}(\mathsf{EvalRotate}(c_{neg},\, r),\, m_{aux})$
\State $c_{new} \leftarrow \mathsf{EvalAdd}(c_{payload},\, c_{aux})$
\State \Return $c_{new}$
\end{algorithmic}
\end{algorithm}

All steps are performed homomorphically, ensuring slot occupancy consistency and maintaining statistical privacy. If all payload slots are deleted, the ciphertext is treated as semantically inert. It contributes nothing to aggregate queries, and the auxiliary sum naturally evaluates to zero, guaranteeing that global aggregation remains accurate even in the presence of entirely empty groups.

\subsection{Homomorphic Aggregate and Maintenance}
\label{sec:aggregation}

Building upon the packed layout and the dynamic mutability primitives, Hermes introduces a new mechanism for evaluating analytical queries over encrypted data. This section describes how the system executes global aggregations and how it incrementally maintains the necessary auxiliary statistics to guarantee query correctness.

\subsubsection{Global Aggregation}

Traditional approaches to encrypted aggregation require expensive slot-wise traversal and Galois automorphisms to accumulate values within a ciphertext. Hermes bypasses this bottleneck by leveraging the precomputed local sum embedded in the final slot of each ciphertext.

For global analytical queries such as \texttt{SELECT SUM(attr)}, the database engine does not need to unpack or rotate the individual data slots. Instead, the query planner translates the aggregation into a single ciphertext level addition. Given a collection of ciphertexts corresponding to a queried table or group, Hermes executes a homomorphic addition across all ciphertexts. Because the BFV scheme operates in a SIMD fashion, this single addition simultaneously aggregates the payload slots and the auxiliary sum slots. The database client then decrypts the aggregated result and strictly reads the value at the final slot $n-1$. This transforms an operation that typically requires logarithmic rotation depth into a constant time ciphertext addition, reducing the latency of encrypted analytics. 

\subsubsection{Incremental Maintenance}

To ensure the correctness of aggregation queries over mutable data, the embedded local sum must remain perfectly synchronized with the active payload slots. Whenever an insertion or deletion occurs, Hermes performs a lightweight incremental update on the auxiliary slot rather than recomputing the sum from scratch.

As demonstrated in the mutability primitives, the inserted or deleted value $v$ is encrypted client-side prior to transmission, 
arriving at the server as a fresh ciphertext $c_v \leftarrow \mathsf{Enc}(v)$.
To update the auxiliary slot, Hermes rotates $c_v$ by an offset $r = (n-1) - i$ to align the encrypted value with the reserved slot at index $n-1$, 
then applies a one-hot plaintext mask $m_{\mathit{aux}}$ to zero all other slots, 
producing $c_{\mathit{aux}} = \mathsf{EvalMult}(\mathsf{EvalRotate}(c_v, r),\, m_{\mathit{aux}})$.
A single ciphertext-ciphertext addition $\mathsf{EvalAdd}(c,\, c_{\mathit{aux}})$ then increments the local sum to reflect the newly ingested record.
Conversely, for a deletion, Hermes applies $\mathsf{EvalNegate}$ to $c_v$ before the same rotate-mask-add pipeline, 
decrementing the local sum by the encrypted deleted value.
Both updates require no re-encryption or multi-level circuit evaluations, 
and incur negligible noise growth under the BFV
parameter configuration described in Section~\ref{sec:noise_derivation}.

Algorithm~\ref{alg:aggregation} formalizes the dual mechanism of global aggregation and incremental maintenance, illustrating how Hermes bypasses traditional computational bottlenecks. The procedure is divided into two distinct operational phases. In the first phase, the system evaluates global analytical queries by accumulating a collection of ciphertexts $c_1$ through $c_k$. Instead of unpacking and rotating individual data slots, Hermes iteratively applies homomorphic addition across the entire ciphertext collection. Because the precomputed local sums are deterministically aligned at the final auxiliary slot of each ciphertext, this SIMD operation natively computes the global aggregate using a single arithmetic evaluation per ciphertext. 
The second phase details the lightweight maintenance logic required when the underlying database tuple mutates.
To keep the auxiliary statistic synchronized with the active payload,
the system rotates the encrypted update $c_v \leftarrow \mathsf{Enc}(v)$ by offset $r = (n-1) - i$ to align it with the reserved slot at index $n-1$,
then applies a plaintext selection mask $m_{\mathit{aux}}$ with $1$ at index $n-1$ and $0$s elsewhere, 
producing $c_{\mathit{aux}} = \mathsf{EvalMult}(\mathsf{EvalRotate}(c_v, r),\, m_{\mathit{aux}})$.
A single ciphertext-ciphertext addition $\mathsf{EvalAdd}(c,\, c_{\mathit{aux}})$ then updates the local accumulator without any plaintext exposure or multi-level circuit evaluation.

\begin{algorithm}[t]
\caption{Homomorphic Aggregate and Maintenance}
\label{alg:aggregation}
\begin{algorithmic}[1]
\Require Ciphertexts $c_1, \dots, c_k$, Target $c$, Ciphertext $c_v \gets \mathsf{Enc}(v)$, Target index $i$, Capacity $n$
\Ensure Aggregated result $c_{agg}$, Maintained ciphertext $c_{new}$

\Statex \textit{// Part 1: Accelerated Global Summation}
\State $c_{agg} \gets c_1$
\State \textit{for} each subsequent ciphertext $c_j$ ($j=2 \dots k$) \textit{do}
\State \hspace{1em} $c_{agg} \gets \mathsf{EvalAdd}(c_{agg}, c_j)$ \Comment{Aggregate of auxiliary slots}
\State \textit{end for}

\Statex \textit{// Part 2: Incremental Maintenance of Local Sum}
\State $m_{aux} \gets$ vector with 1 at index $n-1$ and 0s elsewhere
\State $r \gets (n-1) - i$ \Comment{Calculate rotation offset to auxiliary slot}
\State $c_{aux} \gets \mathsf{EvalMult}(\mathsf{EvalRotate}(c_v, r), m_{aux})$
\State $c_{new} \gets \mathsf{EvalAdd}(c, c_{aux})$ \Comment{Ciphertext-ciphertext addition}

\State \Return $c_{agg}$ and $c_{new}$
\end{algorithmic}
\end{algorithm}

\section{Security and Performance Analysis}
\label{sec:analysis}

\subsection{Security Model and Assumptions}
\label{sec:security_model}

We analyze the security of Hermes under the standard indistinguishability under chosen-plaintext attack (IND-CPA) model, instantiated using BFV~\cite{FV2012}. We assume a passive, honest-but-curious adversary with full access to encrypted database contents, including ciphertexts before and after updates, as well as query results. The adversary may observe ciphertext structure, slot layout, and numerical patterns in decrypted outputs but is assumed to follow protocol without actively deviating from the query logic. 
All mutation operands, including inserted and deleted values, are encrypted by the client prior to transmission using the same BFV public key.
The server therefore processes only ciphertexts and public plaintext masks, and never observes any plaintext payload values during insertion or deletion.
However, the server does not record low-level timing, memory access, or power traces: Preventing side-channel attacks is beyond the scope.

We characterize the security guarantees of Hermes through the following invariants, which together define its leakage profile:

\begin{enumerate}
  \item \textit{Indistinguishability of Ciphertexts:} For any two payload vectors $v$ and $v'$ of the same length, their packed ciphertexts $c = \texttt{Encrypt}(v)$ and $c' = \texttt{Encrypt}(v')$ are computationally indistinguishable under the IND-CPA security of BFV.
  
  \item \textit{Structural Profile:} For an insertion or deletion targeting logical slot $i$ within group $g$, the server observes the plaintext position mask $\mathbf{m}^{(i)}$, from which the intra-group slot index $i$ is deducible; the updated value itself remains encrypted and is never exposed.
  
  \item \textit{Aggregation Consistency:} Global aggregation over auxiliary slots (e.g., encrypted `SUM') reveals no more information than the final aggregate result; the per-ciphertext local sum values are protected under the encryption scheme.

  \item \textit{No Adaptive Leakage:} Hermes does not reveal any intermediate plaintexts or randomness that could be used to link ciphertexts over time. All operations are performed homomorphically and reuse no client secrets post-encryption.
\end{enumerate}

The proposed Hermes data model and update interface preserve these properties without relying on secure multi-party computation (MPC), secure enclaves, or trusted client-side computation. All operations, such as packing, insert/delete, and aggregation, are executed on encrypted data within the untrusted database server.

This model captures realistic cloud deployment scenarios where users outsource encrypted databases to an untrusted cloud provider but wish to retain strong confidentiality guarantees under standard cryptographic assumptions (e.g., Ring-LWE hardness). We note that this threat model is stronger than many property-preserving encryption (PPE) systems, which often leak order, frequency, or access pattern by design. However, like any packed FHE design that supports in-place updates, Hermes inevitably reveals the intra-group slot position via the public plaintext mask used in shift-and-mask operations. We consider this structural profile an acceptable and deliberate trade-off to achieve both practical performance and true mutability, 
which have not been simultaneously supported by prior FHE database systems.

\subsection{IND-CPA Security}

The semantic security of Hermes relies on the assumption that the underlying BFV scheme remains IND-CPA secure under all ciphertext manipulations performed in the system. This includes ciphertext packing, masked updates, homomorphic insertion and deletion, and the embedding of auxiliary local statistics such as the group sum. 
We now argue that the structural profile described in Section~\ref{sec:security_model} does not compromise the IND-CPA security of payload values.
The formal proof is based on the following two observations.

First, although Hermes embeds deterministic metadata (e.g., local sums) into specific slots, the entire ciphertext is encrypted using randomized BFV encryption over the full plaintext vector. IND-CPA security guarantees that the ciphertext remains indistinguishable even if some slots are derived from others, so long as the message vector is encrypted as a whole with fresh randomness. In particular, the inclusion of auxiliary statistics does not compromise the semantic security of any individual slot.

Second, Hermes performs all slot-level updates, such as insertion and deletion, purely through homomorphic operations. These updates use masking, shifting, and slot-wise additions, but never invoke decryption, re-encryption, or key switching. No new randomness is injected, and no plaintext values are revealed or reconstructed during runtime. As a result, each modified ciphertext maintains the same IND-CPA security level as its original version with respect to payload confidentiality.

We start by recalling the notion of IND-CPA security.
\begin{definition}[IND-CPA Security]
Let $\mathsf{Enc}_\mathsf{pk}(m; r)$ denote the encryption of a plaintext $m$ under public key $\mathsf{pk}$ and randomness $r$. A scheme is IND-CPA secure if no probabilistic polynomial-time (PPT) adversary $\mathcal{A}$ can distinguish between encryptions of $m_0$ and $m_1$ for any chosen pair of equal-length plaintexts, even given arbitrary access to the encryption oracle. Formally, for any PPT adversary $\mathcal{A}$, the advantage
\[
\text{Adv}_\mathcal{A}^{\text{IND-CPA}} = \left| \Pr[\mathcal{A}(\mathsf{Enc}(m_0)) = 1] - \Pr[\mathcal{A}(\mathsf{Enc}(m_1)) = 1] \right|
\]
is negligible in the security parameter $\lambda$.
\end{definition}

\begin{lemma}
Let $\mathcal{C}$ be the set of ciphertexts generated and manipulated by Hermes. Then under the assumption that the base FHE scheme (BFV) is IND-CPA secure, the ciphertext $c \in \mathcal{C}$ remains IND-CPA secure throughout all operations in Hermes.
\end{lemma}

\begin{proof}[Proof Sketch]
The proof relies on a reduction to the underlying IND-CPA security of the BFV scheme. Since Hermes constructs its packed data model using standard plaintext encoding and standard homomorphic evaluations, any adversary capable of distinguishing between two Hermes ciphertexts could be leveraged to break the semantic security of the base BFV encryption. 

Furthermore, the deterministic inclusion of auxiliary local sums and the application of public plaintext masks during updates do not compromise the IND-CPA security of payload values. All slot modifications are performed strictly within the encrypted domain without relying on secret-dependent control flows.

The full proof can be found in Appendix~\ref{appendix:proofs}, where we construct a simulator that interacts with the IND-CPA challenger for BFV and demonstrates that any advantage in distinguishing Hermes ciphertexts translates to an advantage against the base scheme.
\end{proof}

An important subtlety in Hermes is the deterministic embedding of auxiliary data into a fixed slot. While this creates a structural invariant, it does not violate IND-CPA since the data being embedded is functionally independent on other encrypted slots. From a cryptographic perspective, this corresponds to publishing a function $f(v_0, \dots, v_{n-2}) = s$ along with the encryption of the vector, which is a standard practice in FHE-based secure computation.
Furthermore, all masking and slot operations are purely homomorphic and do not involve conditional branching or secret-dependent control flow. No part of the computation leaks information via side channels, timing, or slot access patterns. 

\subsection{Correctness of Updates and Aggregates}

We now formally prove that Hermes \emph{correctly} maintains the semantics of encrypted vector updates and aggregate computations in the presence of insertions and deletions. 

\begin{theorem}
  Hermes guarantees the following three properties for any sequence of homomorphic insertions, deletions, and global aggregations over packed ciphertexts:
  \begin{enumerate}
    \item The logical payload vector within each ciphertext remains consistent with the intended insertions and deletions, preserving the correct order and values of active slots.
    \item The auxiliary sum slot accurately reflects the current sum of all valid payload slots after any sequence of updates, ensuring that local statistics remain correct.
    \item The result of encrypted aggregation (e.g., $`SUM'$) over multiple ciphertexts yields the correct aggregate value corresponding to the underlying plaintexts, even in the presence of concurrent updates.
  \end{enumerate}
\end{theorem}

\begin{proof}[Proof Sketch]
The correctness of the Hermes data model is proven by algebraically tracing the homomorphic slot manipulations and verifying them against the expected plaintext operations. For insertions and deletions, we demonstrate that the combination of Galois shifting and zeroing masks perfectly preserves the structural order of the active payload slots while maintaining the strict capacity invariants. 

Similarly, the auxiliary local sum is incrementally maintained via homomorphic scalar addition or subtraction, ensuring perfect synchronization with the payload. When global aggregation is evaluated, the homomorphic addition across the aligned auxiliary slots naturally yields the exact cumulative sum of all underlying tuples. A step-by-step formal expansion is available in the extended technical report.

The full proof can be found in Appendix~\ref{appendix:proofs}, where we provide a rigorous algebraic verification of each property, demonstrating that the homomorphic operations in Hermes faithfully implement the intended database semantics under encryption.
\end{proof}

\subsection{Computational Complexity}

The primary advantage of Hermes lies in its reduction of the computational cost of aggregation queries. Table~\ref{tab:complexity} provides a quantitative comparison of Hermes against scalar FHE and standard packed FHE baseline.
This asymptotic improvement is the theoretical foundation for the significant speedups observed in our experimental evaluation. By shifting the complexity from rotation-heavy query time to lightweight update time, Hermes effectively bridges the gap between database performance and cryptographic security.

\begin{table}[t]
\centering
\caption{Computational Complexity Comparison}
\label{tab:complexity}
\begin{tabular}{lccc}
\hline
\text{Operation} & \text{Scalar FHE} & \text{Standard Packed} & \text{Hermes} \\ \hline
\text{Global Sum} & $O(n) \cdot \text{Add}$ & $O(\log n) \cdot \text{Rot}$ & $O(1) \cdot \text{Add}$ \\
\text{Insertion} & $O(1) \cdot \text{Enc}$ & $O(1) \cdot \text{Rot}$ & $O(1) \cdot \text{Rot}$ \\
\text{Deletion} & $O(1) \cdot \text{Update}$ & $O(1) \cdot \text{Rot}$ & $O(1) \cdot \text{Rot}$ \\
\text{Storage} & $O(n) \cdot \text{CT}$ & $O(1) \cdot \text{CT}$ & $O(1) \cdot \text{CT}$ \\ \hline
\end{tabular}
\end{table}

\subsection{Noise Budget}
\label{sec:noise_derivation}

To justify the sustainability of Hermes updates, we provide a formal analysis of noise growth within the BFV encryption context. 
Let $V$ denote the noise variance of a ciphertext. 
Each insertion or deletion primitive consists of one ciphertext-plaintext multiplication followed 
by one Galois rotation. 
For a ciphertext-plaintext multiplication, 
the noise growth is bounded by $V_{new} \leq t \cdot \text{poly}(N) \cdot 
V_{old}$, where $t = 65537$ and $N = 2^{14}$. 
Because we exclusively use plaintext masks rather than ciphertext-ciphertext multiplications, 
noise accumulates linearly rather than quadratically per operation.

We empirically measured the per-operation noise consumption using OpenFHE's built-in noise estimator across 200 sequential update trials. 
Under our parameter configuration ($d=2$, $t=65537$, $N=2^{14}$), 
the initial noise budget is approximately 60 bits. 
Each insertion or deletion consumes on average $\delta_{insert} \approx 1.3$ bits and 
$\delta_{delete} \approx 1.1$ bits respectively, 
yielding the cumulative bound:
\[
\log_2(V_{total}) \approx \log_2(V_{initial}) + k \cdot \bar{\delta},
\]
where $\bar{\delta}$ is the average per-operation noise growth. 
When mutation operands are supplied as fresh ciphertexts rather than plaintext scalars, 
the auxiliary slot update applies a ciphertext-ciphertext addition instead of a plaintext-ciphertext addition.
Empirically, this increases the per-operation noise consumption by at most $0.2$ bits under our parameter configuration ($d = 2$, $t = 65537$, $N = 2^{14}$),
leaving the cumulative bound and the \mbox{45-update} threshold unchanged.

While this 45-update threshold represents a theoretical bound, it aligns with the operational reality of analytical database systems. In typical OLAP workloads, data aggregations are evaluated on a daily or weekly basis. It is exceptionally rare for a single encrypted analytical record to undergo 45 independent updates within such a short temporal window. To ensure long-term sustainability, system administrators can easily schedule routine ciphertext repackaging tasks during off-peak hours. This background maintenance refreshes the noise budget completely without introducing any latency to daytime query executions. Looking forward, recent advancements in reliable non-leveled FHE optimizations \cite{bchen_www26} present a promising architectural alternative. Transitioning to a non-leveled paradigm could theoretically eliminate the noise budget constraint entirely through continuous homomorphic bootstrapping. However, given the substantial computational overhead still associated with bootstrapping operations, Hermes currently retains its leveled design to prioritize sub-millisecond query latency and high-throughput analytical performance, leaving non-leveled integration as a direction for future system iterations. 

\section{System Implementation}
\label{sec:implementation}

We implement \textsc{Hermes} as a suite of MySQL loadable functions written in C++ with OpenFHE v1.2.4. The system consists of modular components for slot-wise packing, homomorphic insertion and deletion, local-sum-aware aggregation, and slot-selective decryption, all of which are executed natively within the MySQL runtime.

As of April 2026, the codebase comprises 5,553 lines of code. All components are implemented in C++ and compiled into dynamic libraries registered as MySQL loadable functions. Our minimalist integration strategy avoids any modification to the MySQL source code or query engine internals. The full source code is publicly available on Github. Experiments and development were partially supported by the NSF-funded Chameleon Cloud infrastructure~\cite{chameleoncloud}. 

More implementation details can be found in Appendix~\ref{appendix:implementation}.

\subsection{Architecture and Integration}

Each cryptographic operation is compiled as a C++ function into a shared object file, which is dynamically linked into the MySQL server process. This mechanism preserves full compatibility with unmodified MySQL 8.x deployments, requiring no server side modifications or query proxies. Each plugin function adheres strictly to the MySQL UDF lifecycle. The \texttt{\_init} function validates argument counts and types while allocating necessary internal states. The core execution resides in \texttt{\_func}, which runs per row during query evaluation. Finally, the optional \texttt{\_deinit} function releases allocated memory and handles. 

To bridge MySQL data types and OpenFHE plaintexts, all inputs undergo explicit parsing across the foreign function interface (FFI) boundary. MySQL UDF arguments arrive as raw C pointers tagged with type enumerations, such as \texttt{INT\_RESULT} or \texttt{STRING\_RESULT}. Hermes implements strict casting logic to convert these inputs into 64-bit signed integers, utilizing \texttt{std::istringstream} when numerical data is passed as string types. These scalar values are subsequently passed to the OpenFHE \texttt{MakePackedPlaintext} API.

\subsection{Ciphertext Management}

Instead of serializing ciphertexts into strings or blobs for every operation, ciphertext objects are allocated on the heap and remain within the memory space of the MySQL process. For diagnostic and linkage purposes, Hermes returns a string-encoded metadata footprint of the ciphertext rather than the serialized payload. This diagnostic string takes the format of \texttt{0xADDR (v, size=s)}, where \texttt{ADDR} is the hexadecimal memory address of the ciphertext object, $v \in \mathbb{Z}$ is a verified decrypted payload used for internal sanity checks, and $s$ represents the object footprint in bytes.

This memory-centric layout minimizes I/O overhead during query execution. When operations like \texttt{HERMES\_PACK\_CONVERT()} are invoked in a query, the plugin parses the string to recover the memory address, safely casts it back to a \texttt{Ciphertext<DCRTPoly>} pointer, and performs the requested homomorphic evaluation directly in memory. This pointer-based memory lifecycle ensures that the heavy cryptographic objects never cross the FFI boundary, keeping the SQL engine lightweight and avoiding deserialization.

\subsection{Context and Key Management}

To ensure consistent cryptographic behavior across all functions, Hermes constructs a shared context during the initialization phase. This context is generated using the OpenFHE \texttt{GenCryptoContext} API. We parameterize the environment with a plaintext modulus $t = 65537$, a ring dimension $N = 2^{14}$, and a multiplicative depth $d = 2$. These parameters are selected to balance computational latency with a sufficient noise budget for query evaluations.

Once the core context is established in memory, we explicitly enable the required OpenFHE operational modes, including Public Key Encryption (PKE), Leveled Somewhat Homomorphic Encryption (SHE), and Advanced SHE to unlock evaluation keys. Key generation occurs once at system startup. \texttt{KeyGen()} is invoked to produce the public and secret keys. Subsequently, Hermes generates the necessary auxiliary cryptographic materials. This involves calling \texttt{EvalMultKeyGen()} to create relinearization keys and \texttt{EvalSumKeyGen()} along with \texttt{EvalAtIndex()} to generate the Galois keys required for slot rotation algorithms. All cryptographic keys and the evaluation context are encapsulated within static C++ global variables protected by thread-safe initialization flags.

\section{Evaluation}
\label{sec:evaluation}

An example list of Hermes APIs is provided in Appendix~\ref{appendix:api}. 

\subsection{Experimental Setup}
\label{sec:datasets}

\paragraph{Platform}
All experiments are conducted on a Dell PowerEdge R6525 server provisioned via the CHI@UC Chameleon Cloud~\cite{chameleoncloud}. The compute node is equipped with dual AMD EPYC 7763 processors operating at 2.45GHz, providing a total of 128 physical cores and 256 hardware threads, alongside 512GB of memory. The system runs on Ubuntu 24.04 LTS and utilizes MySQL version 8.0.42 as the underlying database engine. The homomorphic cryptographic operations are implemented using OpenFHE version 1.2.4 configured with the BFV scheme. For all evaluations, we establish the encryption context using a default ring dimension of $N=2^{14}$ and a plaintext modulus of $t=2^{16}$, which yields exactly $n=8192$ plaintext slots per ciphertext for vectorized execution.

\paragraph{Datasets}
We evaluate Hermes on the standard TPC-H benchmark and three real-world datasets that represent diverse application domains and data distributions. 
\begin{itemize}
  \item The \emph{TPC-H benchmark}~\cite{tpch3} is a widely adopted standard for evaluating analytical query performance. We focus on the \texttt{lineitem} and \texttt{orders} tables, which contain 6 million and 1.5 million tuples respectively at scale factor 1. These tables are rich in numerical attributes suitable for homomorphic aggregation.
  \item The \textit{COVID-19} dataset~\cite{covid19data} contains 341 daily records of pandemic-related statistics in the U.S., such as hospitalizations and case counts. 
  \item The \textit{hg38} dataset~\cite{hg_data} corresponds to genomic annotations in the human reference genome, consisting of over 34,000 entries such as transcription start sites and exon counts. 
  \item The \textit{Bitcoin} dataset~\cite{bitcoin_trade} records cryptocurrency trade volumes on a 3-day interval from 2013 to 2022. Since raw Bitcoin trade values are extremely large, we apply a rescaling factor of $1/24$ to obtain minute-level approximations before encryption.
\end{itemize}

\paragraph{Baselines}
To evaluate the architectural and algorithmic advantages of Hermes, we establish four baselines. We partition these baselines into end-to-end system macrobenchmarks within the database engine and isolated cryptographic microbenchmarks.

For the end-to-end evaluations within MySQL, we compare Hermes against a Singular FHE approach and a Plaintext baseline. The Singular FHE baseline represents a naive integration where each database tuple is encrypted into an independent ciphertext without SIMD batching. This configuration highlights the computational and storage overheads of unpaced encryption and justifies our vectorized data model. The Plaintext baseline operates on completely unencrypted data using standard SQL queries. It serves as the absolute performance ceiling to quantify the exact system-level penalty introduced by our cryptographic layer.

To isolate our algorithmic improvements from the inherent execution overhead of the database engine, we implement two additional baselines as standalone C++ microbenchmarks. The first is a Standard Packed FHE baseline, which utilizes the same multi-slot ciphertext layout as Hermes but relies on the conventional sequence of logarithmic Galois rotations and masking operations to compute aggregate sums. This comparison isolates the latency reduction achieved specifically by our auxiliary local sum embedding technique. The second microbenchmark employs the Paillier cryptosystem~\cite{paillier1999}, representing the traditional Partially Homomorphic Encryption (PHE) standard for secure database aggregations. Because Paillier supports only scalar additive homomorphism, this evaluation demonstrates how the vectorized parallelism of modern FHE outpaces legacy scalar schemes in analytical throughput.

While several homomorphic prototypes such as HE3DB~\cite{10.1145/3576915.3616608} and ArcEDB~\cite{10.1145/3658644.3670384} exist, we exclude them from empirical comparison for two reasons. First, these works function as client-side standalone scripts rather than being integrated into the execution pipeline of a production-grade DBMS. Second, Hermes prioritizes strict IND-CPA security without access pattern leakage, whereas those alternatives rely on interactive protocols or property-preserving encryption that inherently weaken the threat model. Thus, we benchmark against Intel IPCL~\cite{ipcl_github} and OpenFHE~\cite{openfhe} to establish a baseline using identical cryptographic security levels.

\paragraph{Objectives and Metrics}
We structure experiments around four research questions (RQs):
\begin{itemize}
    \item RQ1: How does Hermes perform across standard analytical workloads and maintenance operations when evaluated against a production-grade benchmark like TPC-H? We address this in Section~\ref{sec:eval_tpch}.
    \item RQ2: How does the vectorized packing data model improve initial data ingestion and encryption throughput compared to singular tuple encryption? We address this in Section~\ref{sec:eval_encryption}.
    \item RQ3: What is the computational latency of performing secure in-place modifications such as record insertions and deletions over packed ciphertexts? We explore this in Section~\ref{sec:eval_update}.
    \item RQ4: How does the auxiliary local-sum embedding technique perform against traditional cryptographic aggregation queries in isolated microbenchmarks? We answer this in Section~\ref{sec:eval_microbenchmarks}.
    \item RQ5: How does the logical group capacity influence the overall execution latency across different database operators? We analyze this in Section~\ref{sec:eval_scalability}.
    \item RQ6: To what extent does the vectorized packing model mitigate physical storage inflation across diverse datasets? We evaluate this in Section~\ref{sec:eval_storage}.
\end{itemize}

\subsection{TPC-H Benchmark}
\label{sec:eval_tpch}

To evaluate the robustness and scalability of Hermes in standard analytical scenarios, we conduct a comprehensive evaluation using the TPC-H benchmark. Our experiments focus on two representative tables: the \texttt{lineitem} table for analytical queries and the \texttt{orders} table for data maintenance operations. We vary the tuple count within a single logical group from 1,000 to 15,000 to analyze the scalability of three distinct workloads: (1) simplified Q1 aggregation (\texttt{SUM}), (2) record insertion, and (3) record deletion.

\begin{figure}[!t]
\centering
\includegraphics[width=\columnwidth]{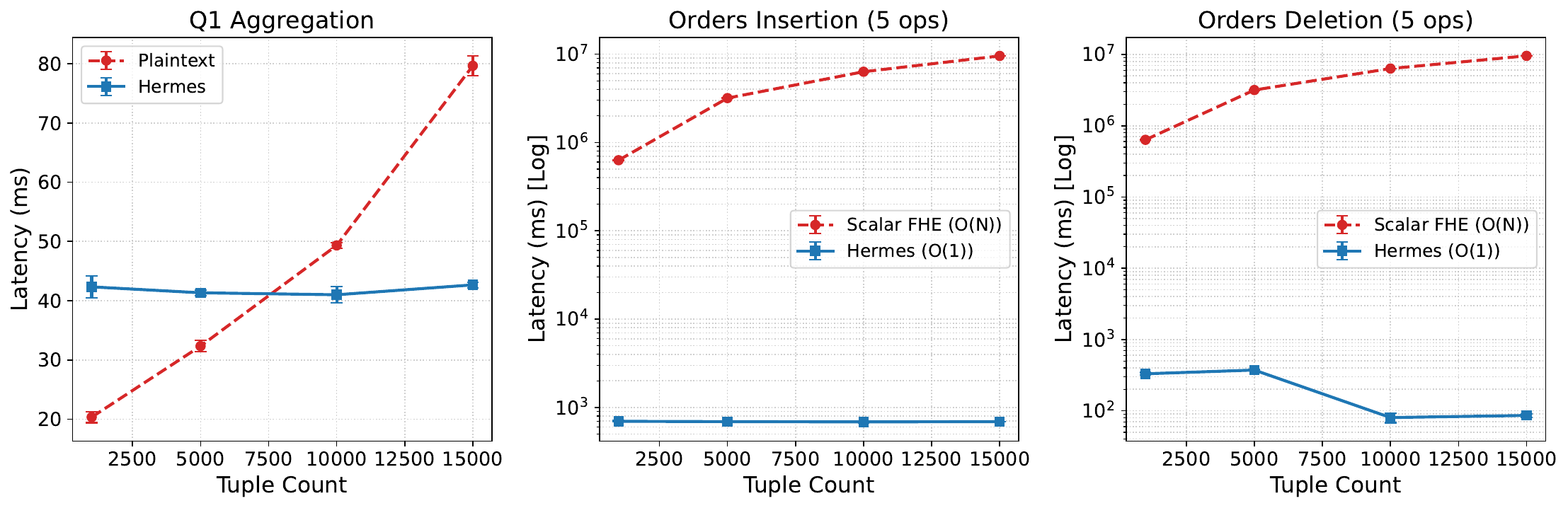}
\caption{Three workloads on TPC-H \texttt{lineitem} and \texttt{orders} tables across varying tuple counts. 
}
\label{fig:tpch_eval}
\end{figure}

\paragraph{Workload 1: Q1 Aggregation} 
We first evaluate the performance of a global \texttt{SUM} aggregation on the \texttt{l\_quantity} column of the \texttt{lineitem} table, comparing Hermes against a plaintext MySQL baseline. As shown in Figure \ref{fig:tpch_eval}(a), the latency of native plaintext SQL grows linearly from 20~ms to 80~ms as the tuple count increases, reflecting the O(N) complexity of row-by-row scanning. In contrast, Hermes exhibits near-constant latency ($\approx$42~ms) across all scales. This is because Hermes leverages precomputed auxiliary sum slots embedded during ingestion, transforming the query into a constant-time ciphertext addition. Notably, Hermes achieves a performance crossover, outperforming plaintext MySQL once the group size exceeds 10,000 tuples.

\paragraph{Workload 2 \& 3: Record Modifications.} 
We evaluate the mutability primitives using the \texttt{o\_totalprice} column in the \texttt{orders} table. We compare Hermes's packed mutability against a Scalar FHE baseline, which must process every record in a group to maintain access pattern privacy (O(N) I/O cost). 
As illustrated in Figures \ref{fig:tpch_eval}(b) and \ref{fig:tpch_eval}(c), the Scalar FHE baseline suffers from a prohibitive linear increase in latency, reaching over 9,000 seconds for 5 operations at the 15,000-tuple scale. Conversely, Hermes maintains almost constant-time execution for both insertions and deletions. By utilizing homomorphic shift-and-mask operations directly on packed ciphertexts, Hermes decouples the cost of record updates from the data volume, achieving orders-of-magnitude speedups compared to conventional scalar deployments.

\subsection{Encryption Throughput}
\label{sec:eval_encryption}

\begin{figure}[!t]
\centering
\includegraphics[width=\columnwidth]{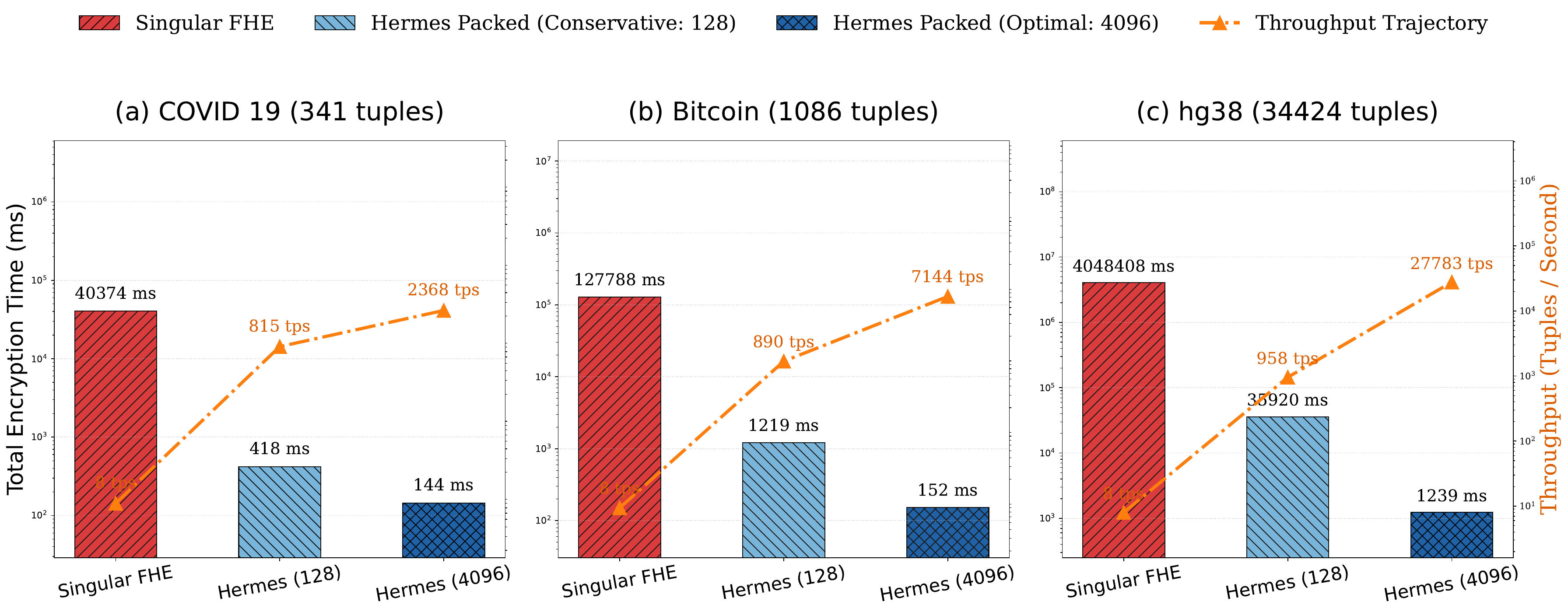}
\caption{Encryption throughput and absolute time comparison across three datasets. 
}
\label{fig:encryption}
\end{figure}

We evaluate the initial data ingestion performance by focusing strictly on the cryptographic overhead. We compare the Hermes vectorized packing model at its boundary capacities against the Singular FHE baseline. Figure~\ref{fig:encryption} presents both the absolute time required to encrypt the complete tables and the normalized throughput measured in tuples per second. The vertical axes utilize a logarithmic scale to properly depict the massive performance gap.

The Singular FHE baseline forces the database to generate an independent ciphertext for every single tuple. As shown in Figure~\ref{fig:encryption}(c), encrypting the hg38 dataset tuple by tuple consumes over 4048 seconds, yielding a slow throughput of merely 8 tuples per second. By contrast, Hermes leverages the SIMD capabilities of the BFV scheme to pack multiple database tuples into a single ciphertext. Even under the most conservative structural configuration of 128 slots, the encryption time for hg38 drops to 35 seconds, boosting the throughput to 958 tuples per second.

When operating at the optimal packing scale of 4096 slots, Hermes reduces the total encryption time for hg38 to 1.2 seconds. The overlaid throughput trajectory highlights this dramatic efficiency gain, revealing a high processing rate of 27783 tuples per second, which translates to a staggering 3472x speedup over the singular encryption baseline. Similar computational boundaries are observed in the Bitcoin and COVID-19 datasets. These measurements confirm that the vectorized data model successfully reduces the fundamental throughput bottleneck of applying homomorphic encryption to outsourced database systems.

\subsection{Update Latency}
\label{sec:eval_update}

We evaluate the system performance for in-place data modifications. We focus on the amortized latency per tuple to accurately reflect the throughput capabilities of the vectorized packing architecture under bulk operation workloads.

\begin{figure}[!t]
\centering
\includegraphics[width=\columnwidth]{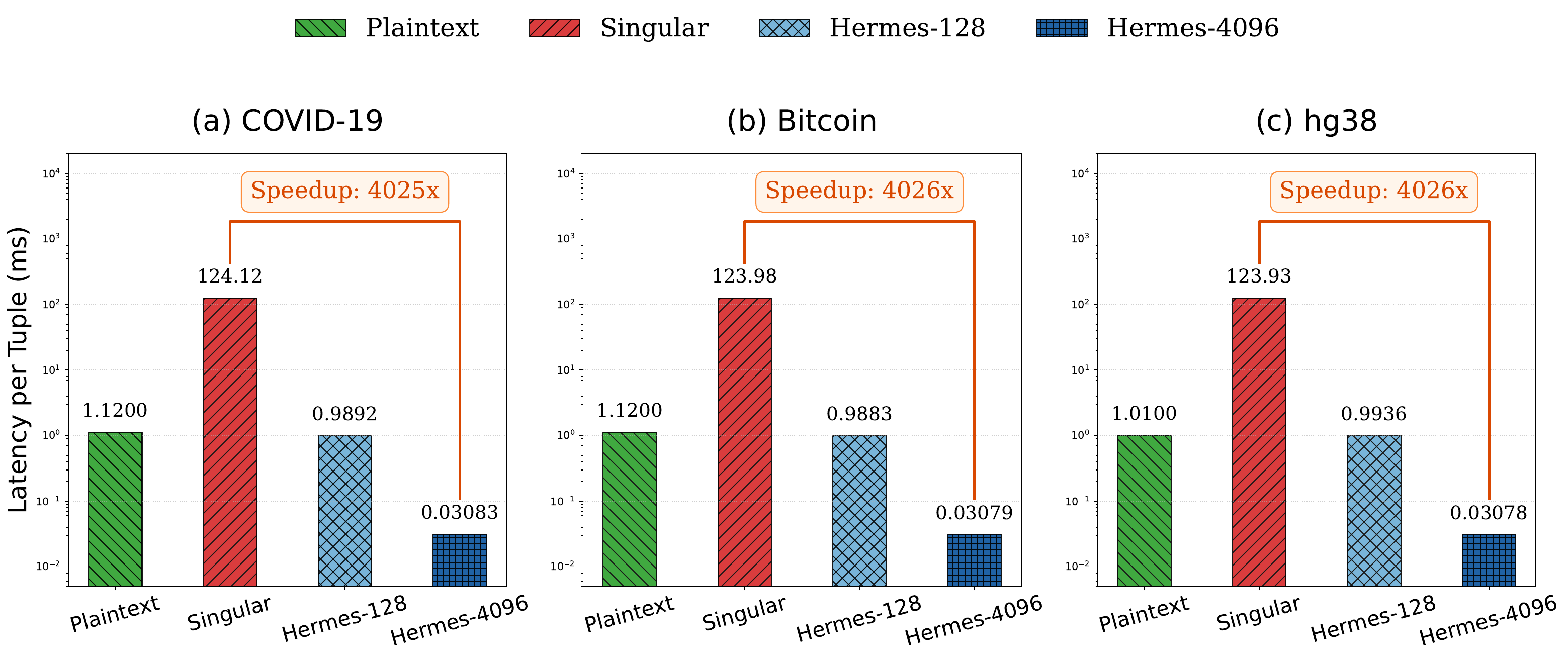}
\caption{Amortized in-place insertion latency per tuple across three evaluation datasets. 
}
\label{fig:insert}
\end{figure}

\begin{figure}[!t]
\centering
\includegraphics[width=\columnwidth]{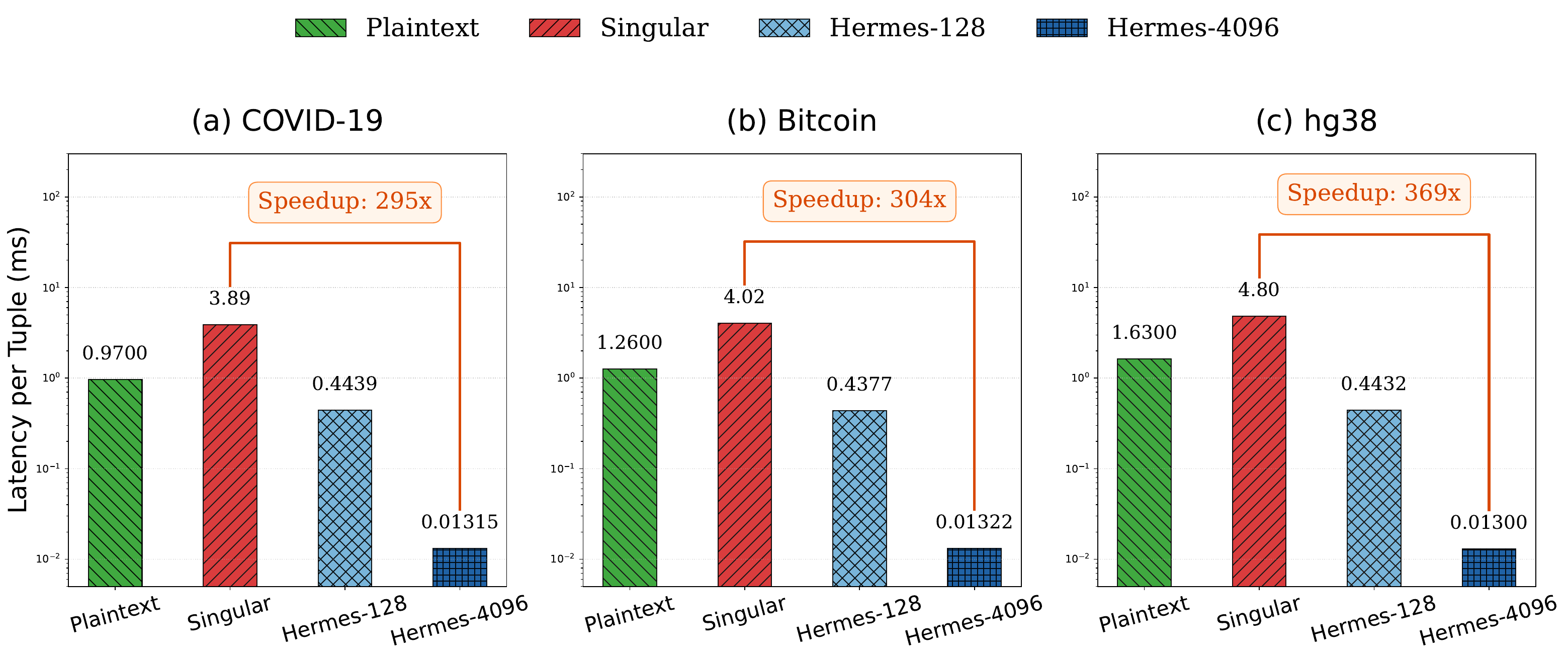}
\caption{Amortized in-place deletion latency per tuple across three evaluation datasets. 
}
\label{fig:remove}
\end{figure}

Figure~\ref{fig:insert} presents the amortized latency for insertion operations. The Singular FHE baseline exhibits a severe bottleneck, requiring approximately 124 milliseconds to insert a single tuple. This delay stems from the mathematical requirement to generate a complete polynomial ciphertext for every isolated record. Hermes leverages the SIMD slots to distribute this cryptographic overhead. Under the optimal configuration of Hermes-4096, the amortized insertion time is reduced to approximately 0.03 milliseconds per tuple. This structural advantage translates to an over 4000x speedup for the Bitcoin dataset and similar gains across the other workloads.

Figure~\ref{fig:remove} illustrates the performance of deletion operations. In the Singular FHE system, a deletion is a direct SQL drop operation requiring merely 4 milliseconds. Hermes cannot physically drop rows within a packed ciphertext and must instead execute homomorphic masking to erase specific slots. Although a single packed deletion takes around 54 milliseconds to compute, amortizing this operation across 4096 slots yields an effective latency of 0.013 milliseconds per tuple. As highlighted by the overlaid metrics, this results in a 369x speedup over the singular approach for the hg38 dataset. Other two datasets show similar trends with speedups around 300x.

These empirical results validate that the SIMD packing design does not degrade the practical update performance. By amortizing the cryptographic polynomial operations across thousands of database tuples, Hermes successfully maintains sub-millisecond effective latencies for both insertions and deletions.

\subsection{Aggregation Queries}
\label{sec:eval_microbenchmarks}

We evaluate the performance of Hermes on aggregation queries. To isolate the intrinsic cryptographic efficiency of our data model from the inherent execution overhead of the MySQL engine, we conduct these evaluations at the core operator level. Although Hermes natively processes SQL aggregations, measuring them end-to-end would introduce parsing and scheduling artifacts that obscure the pure cryptographic throughput. Therefore, we implement this section's benchmarks as standalone C++ programs. This approach allows us to directly compare our rotation-free aggregation technique against traditional cryptographic methods under strict computational isolation.
All benchmarks are executed with $N=2^{14}$ and $t=65537$, yielding 8192 slots.

\begin{figure}[!t]
\centering
\includegraphics[width=\linewidth]{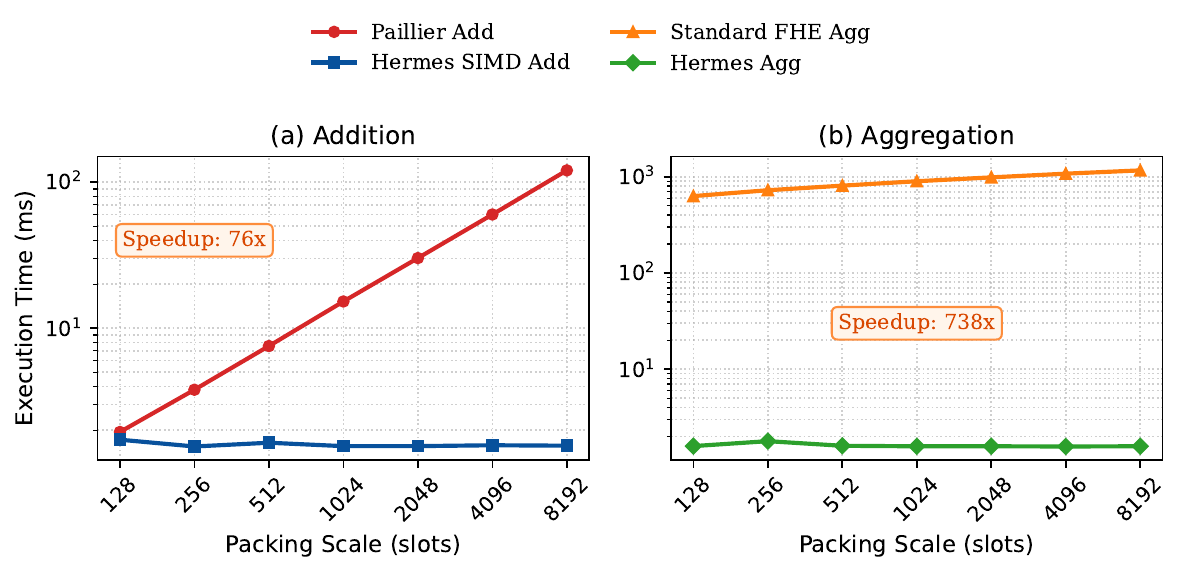}
\caption{Cryptographic query microbenchmarks.
}
\label{fig:microbenchmarks}
\end{figure}

Figure~\ref{fig:microbenchmarks}(a) compares the throughput of encrypted addition between the proposed Hermes data model and the Intel Paillier Cryptosystem Library (IPCL)~\cite{ipcl_github}. While Paillier is a highly optimized partially homomorphic scheme, its scalar nature requires $n$ separate ciphertext additions to process $n$ tuples. As the packing scale increases to 8192 slots, Paillier's latency grows linearly to 120.5 ms. In contrast, Hermes utilizes the SIMD capabilities of BFV to perform a single vector-add operation that processes all 8192 slots simultaneously. This results in a near-constant latency of 1.57 ms, representing a 76x throughput advantage at the maximum scale.

Figure~\ref{fig:microbenchmarks}(b) evaluates the scalability of encrypted aggregation. Standard FHE aggregation strategies rely on a sequence of $\log n$ Galois rotations and relinearizations to sum internal slots. At 8192 slots, this conventional approach consumes over 1165 ms per ciphertext. Hermes fundamentally bypasses this bottleneck by embedding a precomputed local sum into a reserved auxiliary slot during the packing phase. Global aggregation in Hermes is thus reduced to a single ciphertext-level addition over these auxiliary slots. As shown in the results, Hermes completes the same aggregation in 1.58 ms, achieving a 738x speedup over the rotation-based baseline.

These microbenchmarks demonstrate that the performance gains in Hermes are not merely implementation-level optimizations but stem from a structural shift in how homomorphic primitives are applied to database data. By aligning the database layout with the algebraic invariants of SIMD-FHE, Hermes transforms traditionally expensive cryptographic tasks into constant-time operations.

\subsection{Scalability}
\label{sec:eval_scalability}

\begin{figure}[!t]
\centering
\includegraphics[width=\columnwidth]{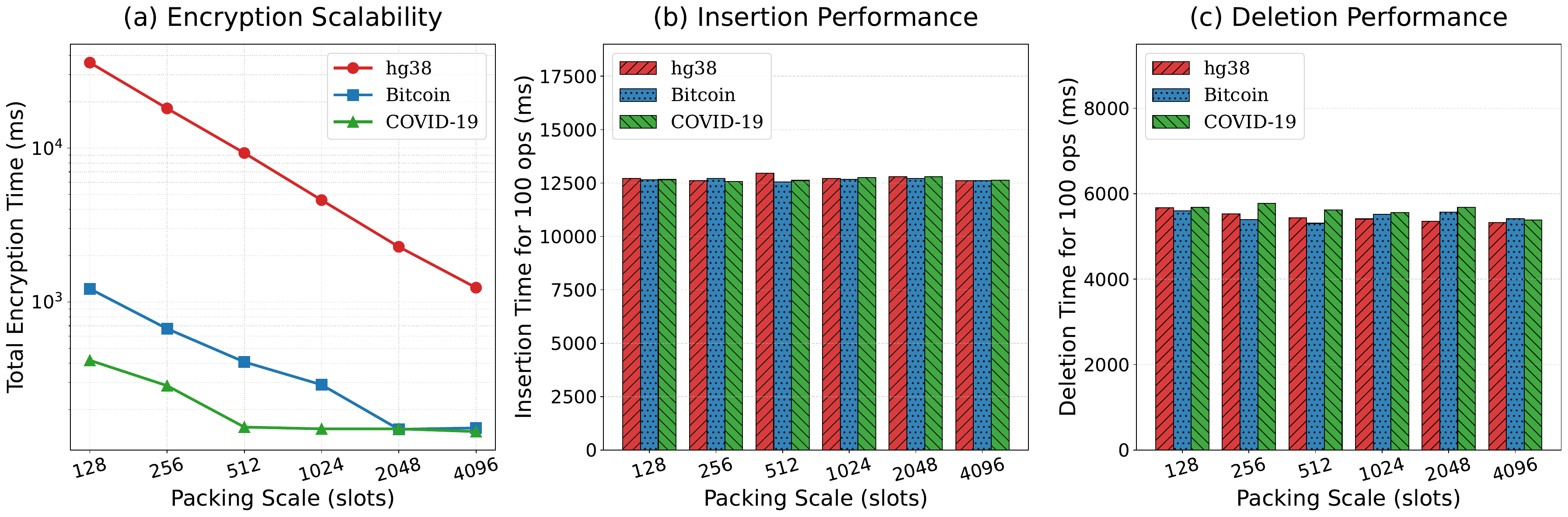}
\caption{Performance scalability of Hermes across varying packing scales from 128 to 4096 slots. 
}
\label{fig:scalability}
\end{figure}

We analyze the system performance across different logical group capacities ranging from 128 to 4096 slots. Figure~\ref{fig:scalability}(a) demonstrates a clear inverse relationship between the packing scale and the encryption time. As the packing capacity doubles, the number of required physical ciphertexts is reduced by half. This leads to an exponential reduction in the overall cryptographic overhead. For example, encrypting the complete hg38 dataset takes over 35 seconds with a pack size of 128 but drops to approximately 1.2 seconds when the scale expands to 4096. This validates Hermes' design that maximizing the utilization of plaintext slots effectively amortizes the initial data ingestion latency.

Conversely, Figure~\ref{fig:scalability}(b) and Figure~\ref{fig:scalability}(c) reveal that the performance of in-place modifications remains entirely decoupled from the packing scale. The time required to perform 100 insertions or 100 deletions stays roughly constant across all scales. Specifically, the batch insertions stabilize at approximately 12.6 seconds, while the corresponding deletions take about 5.4 seconds. 
This constant time complexity is a direct manifestation of the underlying homomorphic mechanics. Because the polynomial ring dimension is fixed at $N=2^{14}$ during the cryptographic context initialization, the computational cost of shifting and masking operations depends solely on the algebraic structure rather than the logical payload. Processing a ciphertext containing 128 active slots requires the exact same CPU cycles as processing one with 4096 active slots. Consequently, system administrators can safely maximize the packing scale to optimize storage footprints and encryption throughput without incurring any latency penalties during subsequent updates.

\subsection{Storage Efficiency}
\label{sec:eval_storage}

We evaluate the physical storage footprint of Hermes compared to the Singular FHE baseline. For the hg38 dataset, which contains 34,424 records, the Singular FHE approach generates one independent ciphertext per tuple. At a ring dimension of $N = 2^{14}$, each BFV ciphertext object occupies approximately 512 KB of memory. Thus, the total storage for the Singular baseline exceeds 17 GB. 

By contrast, Hermes utilizes its SIMD-aware data model to batch 8192 tuples per ciphertext. For the same hg38 dataset, Hermes requires only 5 physical ciphertexts to store the entire logical table. Our empirical measurements show a total storage footprint of only 2.6 MB for Hermes, representing a reduction of over 6,000x in physical storage volume. Table~\ref{tab:storage_summary} provides a summary of the storage footprints across all evaluated datasets. This optimization is a result of slot utilization and the elimination of the massive object-level metadata overhead associated with maintaining tens of thousands of individual FHE objects in the MySQL engine.

\begin{table}[t]
\centering
\caption{Storage Footprint Comparison Across Datasets}
\label{tab:storage_summary}
\begin{tabular}{l c c c}
\hline
Dataset & Singular FHE & Hermes Space & Reduction \\ \hline
COVID-19 & 170.5 MB & 0.5 MB & 341x \\
Bitcoin & 543.0 MB & 0.5 MB & 1086x \\
hg38 & 17.2 GB & 2.6 MB & 6615x \\ \hline
\end{tabular}
\end{table}

\section{Conclusion}
\label{sec:conclusion}

This paper introduces Hermes, a prototype system designed to execute secure and efficient global aggregation queries over homomorphically encrypted data. By incorporating a packed data model with embedded auxiliary aggregates, Hermes achieves constant-time encrypted aggregation while bypassing expensive cryptographic operations. Furthermore, the introduction of mutability primitives allows for dynamic tuple-level insertions and deletions directly on packed ciphertexts via homomorphic shift-and-mask operations. Theoretical analysis confirms that these operations maintain the IND-CPA security guarantees of the underlying BFV scheme while sustaining a practical noise budget.   
Our implementation and evaluation validate that Hermes yields orders-of-magnitude improvements in global query latency, update throughput, and storage efficiency compared to traditional FHE deployments. 

Our future work will explore the extension of Hermes to support more complex query patterns, such as multi-dimensional groupings and table joins. Additionally, we plan to investigate adaptive packing strategies that dynamically adjust the logical group capacity based on workload characteristics to further optimize performance. 

\section*{Acknowledgments} Results presented in this paper were obtained using the Chameleon testbed supported by the National Science Foundation.

\bibliographystyle{ACM-Reference-Format}
\bibliography{ref}

\clearpage
\appendix

\section{Full Proofs}
\label{appendix:proofs}

\begin{lemma}
Let $\mathcal{C}$ be the set of ciphertexts generated and manipulated by Hermes. Then under the assumption that the base FHE scheme (BFV) is IND-CPA secure, the ciphertext $c \in \mathcal{C}$ remains IND-CPA secure throughout all operations in Hermes.
\end{lemma}

\begin{proof}
The proof proceeds by reduction. Suppose there exists an adversary $\mathcal{A}_\text{Hermes}$ that can distinguish two ciphertexts $c_0, c_1 \in \mathcal{C}$ produced under Hermes’ operations with non-negligible advantage. Then we construct an adversary $\mathcal{A}_\text{BFV}$ that breaks IND-CPA security of the underlying BFV scheme.

We consider the following components:
\begin{itemize}
  \item \textbf{Packing}: When packing multiple values into a single ciphertext, we construct a vector $v = (v_0, \dots, v_{n-2}, s)$ where $s = \sum_{i=0}^{n-2} v_i$. The encryption $\mathsf{Enc}(v)$ is indistinguishable from $\mathsf{Enc}(v')$ for any $v'$ of same dimension, as this reduces to IND-CPA security over vector plaintexts in BFV.

  \item \textbf{Auxiliary Slot}: Including a local sum $s$ in the last slot of the plaintext vector does not affect security, since $s$ is deterministically computed from values already encrypted. Its inclusion does not leak new information beyond what is already present in the vector.

  \item \textbf{Insertion}: To insert a plaintext $v$ into a packed ciphertext $c$, we encrypt $v$ into a ciphertext $c_v$, construct a plaintext mask $m$ with $m[i] = 1$, and compute $c' = c + \mathsf{EvalMult}(c_v, m)$. The resulting $c'$ remains IND-CPA secure because it is the sum of IND-CPA secure ciphertexts and homomorphic evaluations thereof. The slot mask $m$ is public and fixed, so no leakage is introduced.

  \item \textbf{Deletion}: Similar reasoning applies. The rotated ciphertext and masked-out tail are both outputs of deterministic homomorphic operations on encrypted data, preserving indistinguishability under the semantic security of BFV.

  \item \textbf{Aggregation}: Aggregation over auxiliary slots only involves ciphertext addition. Since BFV supports additive homomorphism, and all ciphertexts involved are IND-CPA secure, the sum remains indistinguishable.
\end{itemize}

Thus, any distinguishing advantage of $\mathcal{A}_\text{Hermes}$ implies an advantage in distinguishing $\mathsf{Enc}(v)$ vs $\mathsf{Enc}(v')$, contradicting the IND-CPA security of BFV.
\end{proof}

\begin{theorem}
  Hermes guarantees the following three properties for any sequence of homomorphic insertions, deletions, and global aggregations over packed ciphertexts:
  \begin{enumerate}
    \item The logical payload vector within each ciphertext remains consistent with the intended insertions and deletions, preserving the correct order and values of active slots.
    \item The auxiliary sum slot accurately reflects the current sum of all valid payload slots after any sequence of updates, ensuring that local statistics remain correct.
    \item The result of encrypted aggregation (e.g., $`SUM'$) over multiple ciphertexts yields the correct aggregate value corresponding to the underlying plaintexts, even in the presence of concurrent updates.
  \end{enumerate}
\end{theorem}

\begin{proof}

We assume a ciphertext $c \leftarrow \mathsf{Encrypt}([v_0, \dots, v_{n-2}, \sigma])$, where $v_i \in \mathbb{Z}_t$ is the plaintext payload at slot $i$ and $\sigma = \sum_{i=0}^{n-2} v_i$ is the precomputed local sum stored in the auxiliary slot $n-1$.

\paragraph{(1) Correctness of Homomorphic Insertion.}
Let $v_{\text{new}} \in \mathbb{Z}_t$ be a value inserted into position $i$ (with $0 \leq i \leq n-2$). Define the updated plaintext vector as:
$$[v_0', \dots, v_{n-2}'] = [v_0, \dots, v_{i-1}, v_{\text{new}}, v_i, \dots, v_{n-3}].$$
The homomorphic insertion process rotates the suffix 
$$\{v_i, \dots, v_{n-2}\}$$ 
one position to the right using a Galois automorphism, masks the insertion point $i$ with a one-hot vector $m^{(i)} \in \mathbb{Z}_t^n$ such that $m^{(i)}[i] = 1$ and $0$ elsewhere, and adds $v_{\text{new}} \cdot m^{(i)}$ via plaintext multiplication.
Let $c_{\text{ins}} = \mathsf{EvalAdd}(c_{\text{shifted}}, \mathsf{EvalMult}(\mathsf{Encrypt}(v_{\text{new}}), m^{(i)}))$. 
Then, under correct decryption:
$$\mathsf{Decrypt}(c_{\text{ins}})[j] = \begin{cases} v_j & \text{if } j < i \\ v_{\text{new}} & \text{if } j = i \\ v_{j-1} & \text{if } i < j \leq n-2 \\ \sigma + v_{\text{new}} & \text{if } j = n-1 \end{cases},$$
which matches the intended semantics. 
Note that the local sum is homomorphically updated by adding the aligned ciphertext $c_{aux}$, where $c_{aux} = \mathsf{EvalMult}(\mathsf{EvalRotate}(c_v, (n-1)-i), m_{aux})$, such that:
$$\sigma' = \sigma + v_{\text{new}} \in \mathbb{Z}_t$$
is correctly maintained through ciphertext-ciphertext addition on the last slot.

\paragraph{(2) Correctness of Homomorphic Deletion.}
Let $v_i$ be deleted from position $i$. The deletion logic performs a Galois rotation that left-shifts all slots $j > i$, then applies a mask $z \in \mathbb{Z}_t^n$ such that $z[n{-}1] = 0$ and $z[j] = 1$ for $j < n-1$, to zero out the stale tail. The resulting ciphertext $c_{\text{del}}$ decrypts to:
\[
\mathsf{Decrypt}(c_{\text{del}})[j] =
\begin{cases}
v_j & \text{if } j < i \\
v_{j+1} & \text{if } i \leq j \leq n-3 \\
0 & \text{if } j = n-2 \\
\sigma - v_i & \text{if } j = n-1
\end{cases},
\]
where $\sigma$ is decremented by adding the aligned negated ciphertext $c_{\mathit{aux}} = \mathsf{EvalMult}(\mathsf{EvalRotate}(\mathsf{EvalNegate}(c_v),\, (n-1)-i),\, m_{\mathit{aux}})$ via a single ciphertext-ciphertext addition on the auxiliary slot,
yielding $\sigma' = \sigma - v_i \in \mathbb{Z}_t$. The vector semantics remain preserved, and auxiliary statistics remain valid.

\paragraph{(3) Correctness of Global Aggregation.}
Let $c^{(1)}, \dots, c^{(k)}$ be a collection of packed ciphertexts, each of form:

\[
\begin{aligned}
c^{(\ell)} &= \mathsf{Encrypt}\left(
  \left[
    v^{(\ell)}_0,\, v^{(\ell)}_1,\, \dots,\, v^{(\ell)}_{n-2},\, \sigma^{(\ell)}
  \right]
\right), \\
\sigma^{(\ell)} &= \sum_{j=0}^{n-2} v^{(\ell)}_j.
\end{aligned}
\]
Then, the encrypted global sum is computed by:
\[
c_{\text{agg}} = \mathsf{EvalAdd}(c^{(1)}, \dots, c^{(k)}),
\]
followed by slot extraction $\mathsf{Extract}(c_{\text{agg}}, n-1)$ to isolate the final aggregated value. 

The decrypted result satisfies:
\[
\mathsf{Decrypt}(c_{\text{agg}})[n-1] = \sum_{\ell = 1}^{k} \sigma^{(\ell)} = \sum_{\ell = 1}^{k} \sum_{j=0}^{n-2} v_j^{(\ell)},
\]
which matches the true plaintext sum across all tuples.
\end{proof}

\section{Implementation Details}
\label{appendix:implementation}

\subsection{MySQL UDF Lifecycle and Memory Handling}

The integration of Hermes into MySQL relies on the User Defined Function (UDF) interface, which follows a strictly defined execution lifecycle. For each SQL function call, the MySQL engine invokes a sequence of initialization, execution, and deinitialization routines. The \texttt{\_init} function is responsible for pre allocating memory and validating the input arguments provided by the SQL parser. During this phase, Hermes performs strict checks on the number of arguments and their associated types to prevent runtime segmentation faults.

The core computational logic resides in the \texttt{\_func} routine, which is executed for every row in the result set. To avoid the overhead of repeated memory allocations, Hermes utilizes the \texttt{ptr} field in the \texttt{UDF\_INIT} structure to maintain persistent states across row invocations. This is particularly vital for maintaining the cryptographic context and temporary evaluation buffers. Finally, the \texttt{\_deinit} function ensures that all heap allocated objects, including ciphertext pointers and intermediate vectors, are properly released to prevent memory leaks in long running database sessions.

\subsection{Cryptographic Context and Key Initialization}

The security and functionality of Hermes depend on the correct initialization of the OpenFHE environment. We configure a shared encryption context during the plugin load phase to ensure that all functions operate under consistent parameters. The system utilizes the BFV scheme with a ring dimension of $N = 2^{14}$ and a plaintext modulus of $t = 65537$. These engineering choices are optimized to provide sufficient multiplicative depth for complex database queries while maintaining sub millisecond latency for basic additions.

Key management is handled through a singleton pattern to prevent redundant key generation. At startup, the system invokes \texttt{KeyGen()} to produce the primary key pair. Subsequently, we generate specialized evaluation keys required for homomorphic operations. This includes relinearization keys by \texttt{EvalMultKeyGen()} and a set of Galois keys produced through the \texttt{EvalSumKeyGen()} and \texttt{EvalAtIndex()} methods. These keys are stored in volatile memory and protected by mutex locks to ensure thread safe access during concurrent query execution.

\subsection{Deployment and Linker Configurations}

Deploying Hermes on production environments like Ubuntu 24.04 LTS requires careful configuration of the dynamic linker. Since the MySQL daemon runs as a restricted service, it does not naturally inherit the user environment variables. To make the OpenFHE shared libraries visible to the \texttt{mysqld} process, we generate a systemd override file that explicitly sets the \texttt{LD\_LIBRARY\_PATH}. This ensures that the plugin can resolve dependencies like \texttt{libOPENFHEcore.so} at runtime.

Furthermore, all external entry points are wrapped in \texttt{extern `C'} linkage to maintain compatibility with the MySQL symbol resolution logic. The build process is managed via CMake, which handles the static linking of core cryptographic modules while keeping the high level UDF logic relocatable. Diagnostic information and error messages are redirected to the standard error stream, which is automatically captured by the MySQL error log for offline analysis and troubleshooting.

\subsection{Benchmarking and Decryption Harness}

To facilitate performance profiling without the overhead of network communication, Hermes includes a benchmarking harness that allows for local decryption within the server process. This interface is strictly restricted to experimental environments and is not intended for production use. It provides dedicated UDFs like \texttt{HERMES\_DEC\_VECTOR\_SUMMABLE()} to extract scalar results from encrypted aggregates.

The harness utilizes a resident secret key to decrypt result ciphertexts and convert them into standard C++ vectors. We optimize this process by extracting only the requested slot to minimize memory copying. The instrumentation also captures the precise execution time of each homomorphic operator, allowing us to isolate the cryptographic latency from the SQL engine overhead. This detailed profiling was essential for validating the performance gains achieved through the auxiliary sum embedding technique and the vectorized packing model.

\subsection{Foreign Function Interface and Data Parsing}

Bridging the gap between the MySQL C based API and the modern C++17 logic of OpenFHE requires a robust Foreign Function Interface (FFI). All data entering the plugin arrives as raw C pointers within the \texttt{UDF\_ARGS} structure. Hermes implements a custom parsing layer to map these inputs to internal data structures. When numeric data is passed as a string result, the system utilizes the standard \texttt{std::istringstream} object to perform safe conversion to 64 bit signed integers.

For ciphertext management, Hermes avoids the cost of serializing large objects into SQL blobs. Instead, we return a hexadecimal string representation of the memory address. This pointer based approach allows subsequent UDF calls to resolve the ciphertext object by casting the string back to a \texttt{Ciphertext<DCRTPoly>} pointer. To ensure memory safety, every pointer resolution step is accompanied by a validation check to confirm that the address resides within the expected memory segment of the MySQL process.

\subsection{Challenges and Learned Lessons}

Integrating complex cryptographic libraries into a mature database engine reveals several non trivial engineering challenges that are often overlooked in theoretical FHE literature. One primary obstacle is the implicit memory management of the MySQL UDF interface. Specifically, the default 256-byte return buffer for string results is a silent constraint. Exceeding this limit without manual re allocation or careful length tracking leads to silent truncation or memory corruption. Hermes addresses this by implementing a length monitoring layer that bypasses the default buffer when handling large metadata footprints, ensuring that the heavy cryptographic pointers remain intact across the foreign function interface.

Furthermore, maintaining a consistent encryption context across isolated SQL function calls requires a delicate balance between C linkage compatibility and modern C++ idioms. Because UDFs require \texttt{extern `C'} visibility for symbol resolution, the use of RAII and smart pointers for managing OpenFHE objects is restricted at the entry points. This forces the system to rely on manual pointer passing and global heap management, which increases the risk of memory leaks if the MySQL thread handling logic deviates from expected lifecycles. Our experience suggests that the lack of native support for C++ class instances in the UDF interface is a significant hurdle for integrating modern homomorphic libraries.

Finally, we observed that the initialization sequence for homomorphic evaluation keys is extremely sensitive. Failure to generate Galois keys in the precise order, or attempting to regenerate them within a shared library handle, results in undocumented runtime failures that are difficult to trace via standard database logs. Hermes mitigates this by enforcing a singleton-driven initialization protocol that ensures that global cryptographic states are established exactly once during the server startup phase. 

\section{Example APIs}
\label{appendix:api}

Hermes provides a suite of homomorphic encryption operators as native SQL functions. Table~\ref{tab:udf_list} summarizes the core User Defined Functions (UDFs) implemented in our system. Each function is designed to perform a specific cryptographic operation, such as encrypting scalar values, performing homomorphic additions, or managing packed ciphertexts. The source file column indicates where the implementation logic for each UDF can be found within the codebase, allowing developers to easily navigate and understand the underlying mechanics of each operation.

\begin{table}[t]
\centering
\caption{Summary of Hermes homomorphic encryption UDFs and implementation source files.}
\label{tab:udf_list}
\scriptsize
\setlength{\tabcolsep}{2pt}
\begin{tabularx}{\columnwidth}{l X l}
\toprule
UDF Function & Description & Source File \\
\midrule
\texttt{HERMES\_ENC\_SINGULAR\_BFV(v)} & Encrypt scalar integer (slot[0]) & \texttt{singular/udf.cpp} \\
\texttt{HERMES\_DEC\_SINGULAR\_BFV(c)} & Decrypt scalar ciphertext & \texttt{singular/udf.cpp} \\
\texttt{HERMES\_SUM\_BFV(ct)} & Aggregate homomorphically & \texttt{singular/udf.cpp} \\
\texttt{HERMES\_MUL\_BFV(ct1, ct2)} & Multiply two ciphertexts & \texttt{singular/udf.cpp} \\
\addlinespace
\texttt{HERMES\_PACK\_CONVERT(val)} & Pack group values into ciphertext & \texttt{pack/packing.cpp} \\
\texttt{HERMES\_DEC\_VECTOR(ct)} & Decrypt packed vector into CSV & \texttt{pack/packing.cpp} \\
\texttt{HERMES\_PACK\_GROUP\_SUM(v)} & Encrypt per group sum & \texttt{pack/packsum.cpp} \\
\texttt{HERMES\_PACK\_GLOBAL\_SUM(c)} & Add all group ciphertexts & \texttt{pack/packsum.cpp} \\
\addlinespace
\texttt{HERMES\_ENC\_SINGULAR(val)} & Encrypt scalar as BFV & \texttt{pack/packsum.cpp} \\
\texttt{HERMES\_DEC\_SINGULAR(ct)} & Decrypt scalar & \texttt{pack/packsum.cpp} \\
\texttt{HERMES\_PACK\_ADD(c, v, i)} & Insert value in packed ciphertext & \texttt{pack/packupdate.cpp} \\
\texttt{HERMES\_PACK\_RMV(c, i, k)} & Remove slot at index, compact vector & \texttt{pack/packupdate.cpp} \\
\texttt{HERMES\_SUM\_CIPHERS(c1, c2)} & EvalAdd of two ciphertexts & \texttt{pack/packupdate.cpp} \\
\bottomrule
\end{tabularx}
\end{table}

The full implementation of these UDFs includes error handling, memory management, and cryptographic operations that adhere to the security guarantees of the underlying BFV scheme. By providing a comprehensive set of homomorphic primitives directly accessible through SQL, Hermes enables developers to build complex encrypted query logic without needing to manage the intricacies of FHE themselves. The modular design of the source files also allows for easy extension of the UDF library to support additional homomorphic operations or optimizations in the future. For example, the \texttt{HERMES\_PACK\_ADD} function encapsulates the logic for homomorphic insertion into a packed ciphertext, while \texttt{HERMES\_PACK\_RMV} handles the deletion and compaction of slots. These functions abstract away the underlying Galois automorphisms and masking operations, providing a clean interface for SQL queries to manipulate encrypted data efficiently.  

\end{document}